\documentclass[11pt]{article}
\usepackage{jheppub}
\usepackage{mathtools,amssymb}
\usepackage[utf8]{inputenc}
\usepackage{enumitem}
\usepackage{amsmath}
\usepackage{amssymb}
\usepackage{tikz}
\usepackage{simpler-wick}
\usepackage{xcolor}
\usepackage[all]{xy}
\usepackage[export]{adjustbox}
\usepackage{ragged2e}
\usepackage{physics}

\usepackage{babel}
\begin{document}

\global\long\def\aad{(a\tilde{a}+a^{\dagger}\tilde{a}^{\dagger})}%

\global\long\def\ad{{\rm ad}}%

\global\long\def\bij{\langle ij\rangle}%

\global\long\def\df{\coloneqq}%

\global\long\def\bs{b_{\alpha}^{*}}%

\global\long\def\la{\langle}%

\global\long\def\dd{{\rm d}}%

\global\long\def\dg{{\rm {\rm \dot{\gamma}}}}%

\global\long\def\ddt{\frac{{\rm d^{2}}}{{\rm d}t^{2}}}%

\global\long\def\ddg{\nabla_{\dot{\gamma}}}%

\global\long\def\del{\mathcal{\delta}}%

\global\long\def\Del{\Delta}%

\global\long\def\dtau{\frac{\dd^{2}}{\dd\tau^{2}}}%

\global\long\def\ul{U(\Lambda)}%

\global\long\def\udl{U^{\dagger}(\Lambda)}%

\global\long\def\dl{D(\Lambda)}%

\global\long\def\da{\dagger}%

\global\long\def\id{{\rm id}}%

\global\long\def\ml{\mathcal{L}}%

\global\long\def\mm{\mathcal{\mathcal{M}}}%

\global\long\def\mf{\mathcal{\mathcal{F}}}%

\global\long\def\ra{\rangle}%

\global\long\def\kpp{k^{\prime}}%

\global\long\def\lr{\leftrightarrow}%

\global\long\def\lf{\leftrightarrow}%

\global\long\def\ma{\mathcal{A}}%

\global\long\def\mb{\mathcal{B}}%

\global\long\def\md{\mathcal{D}}%

\global\long\def\mbr{\mathbb{R}}%

\global\long\def\mbz{\mathbb{Z}}%

\global\long\def\mh{\mathcal{\mathcal{H}}}%

\global\long\def\mi{\mathcal{\mathcal{I}}}%

\global\long\def\ms{\mathcal{\mathcal{\mathcal{S}}}}%

\global\long\def\mg{\mathcal{\mathcal{G}}}%

\global\long\def\mfa{\mathcal{\mathfrak{a}}}%

\global\long\def\mfb{\mathcal{\mathfrak{b}}}%

\global\long\def\mfb{\mathcal{\mathfrak{b}}}%

\global\long\def\mfg{\mathcal{\mathfrak{g}}}%

\newcommand{\normord}[1]{:\mathrel{#1}:}

\global\long\def\mj{\mathcal{\mathcal{J}}}%

\global\long\def\mk{\mathcal{K}}%

\global\long\def\mmp{\mathcal{\mathcal{P}}}%

\global\long\def\mn{\mathcal{\mathcal{\mathcal{N}}}}%

\global\long\def\mq{\mathcal{\mathcal{Q}}}%

\global\long\def\mo{\mathcal{O}}%

\global\long\def\qq{\mathcal{\mathcal{\mathcal{\quad}}}}%

\global\long\def\ww{\wedge}%

\global\long\def\ka{\kappa}%

\global\long\def\nn{\nabla}%

\global\long\def\nb{\overline{\nabla}}%

\global\long\def\pathint{\langle x_{f},t_{f}|x_{i},t_{i}\rangle}%

\global\long\def\ppp{p^{\prime}}%

\global\long\def\qpp{q^{\prime}}%

\global\long\def\we{\wedge}%

\global\long\def\pp{\prime}%

\global\long\def\sq{\square}%

\global\long\def\vp{\varphi}%

\global\long\def\ti{\widetilde{}}%

\global\long\def\wg{\widetilde{g}}%

\global\long\def\te{\theta}%

\global\long\def\tr{{\rm Tr}}%

\global\long\def\ta{{\rm \widetilde{\alpha}}}%

\global\long\def\sh{{\rm {\rm sh}}}%

\global\long\def\ch{{\rm ch}}%

\global\long\def\Si{{\rm {\rm \Sigma}}}%

\global\long\def\si{{\rm {\rm \sigma}}}%

\global\long\def\sch{{\rm {\rm Sch}}}%

\global\long\def\vol{{\rm {\rm {\rm Vol}}}}%

\global\long\def\reg{{\rm {\rm reg}}}%

\global\long\def\zb{{\rm {\rm |0(\beta)\ra}}}%

\newcommand\JX[1]{\textcolor{purple}{{\it #1}}}
\renewcommand{\op}{\mathrm{op}}
\newcommand{\be}{\begin{equation}}
\newcommand{\ee}{\end{equation}}
\title{Von Neumann Algebras in Double-Scaled SYK}
\author{Jiuci Xu}

\affiliation{Department of Physics, University of California, Santa Barbara, CA 93106, USA}
\emailAdd{Jiuci\_Xu@ucsb.edu}

\justify\abstract{It has been argued that a finite effective temperature
emerges and characterizes the thermal properties of the double-scaled SYK model
in the infinite temperature limit \cite{Lin:2022nss}. Meanwhile, in the static
patch of de Sitter, the maximally entangled state satisfies a KMS condition at
infinite temperature \cite{Witten:2023xze}, suggesting the Type II$_1$ nature of
the observable algebra gravitationally dressed to the observer. In this work, we
analyze the double-scaled algebra generated by chord operators in the
double-scaled SYK model and demonstrate that it exhibits features reflecting
both perspectives. Specifically, we  prove that the algebra is a Type II$_1$
factor, and that the empty state with no open chords defines a tracial state, in
agreement with expectations from \cite{Lin:2022rbf}. We further show that this
state is cyclic and separating for the double-scaled algebra, based on which we
explore its modular structure. We then explore various physical limits of the
theory, drawing connections to JT gravity, the Hilbert space of baby universes,
and Brownian double-scaled SYK.  We also present analytic solutions to the
energy spectrum in both the zero- and one-particle sectors of the left/right
chord Hamiltonian.}

\maketitle

\section{Introduction}
\justify
Over the past few decades, remarkable progress in understanding quantum gravity has been achieved within the framework of the AdS/CFT correspondence  \cite{Maldacena:1997re,Susskind:1993if,Susskind:1994vu,Witten:1998zw,Susskind:1998dq,Giddings:1988cx,Hubeny_2007,Sekino:2008he,Susskind:2014rva,Almheiri_2015,Harlow_2017,Almheiri_2019,Almheiri_2020,penington2020replica,penington2020entanglement,Jafferis_2016,Donnelly:2015hta,Donnelly:2018nbv,Horowitz:1999jd,Dong:2013qoa,Dong:2016fnf,Dong:2017xht,Dong:2018cuv,Dong:2018seb,Giddings:2019hjc,Giddings:2020yes,Kudler-Flam:2023qfl}.
In this setup, gravity admits a description that does not explicitly refer to an
observer. The boundary theory provides a complete and invariant formulation of
the bulk dynamics: physical observables are encoded in correlation functions of
the boundary conformal field theory, independent of any particular reference
frame or observer. In this sense, the dual boundary description offers a
viewpoint in which gravity is effectively turned off, and the notion of an
observer plays no fundamental role.  In contrast, for closed universes or
cosmological spacetimes without asymptotic boundaries, gravity cannot be turned
off. The gravitational degrees of freedom remain dynamical everywhere, and there
is no external vantage point from which to define observables. This leads to
various recent development of invariant formulation of physics that incorporate
observer degrees  of freedom as part of the system, and the description of
gravity in closed universes is expressed in relational terms
\cite{Held:2024rmg,Kaplan:2024xyk}. Equivalently, physical observables must be
gravitationally dressed to ensure diffeomorphism invariance
\cite{PhysRevD.27.2885,Harlow:2023hjb,Giddings:2019hjc}. \footnote{See also
\cite{PhysRevD.27.2885,Marolf_1995,Maldacena_2003}}

Motivated by this perspective, several recent works have explored approaches to de~Sitter gravity that treat the observer as an essential part of the description~\cite{Rahman:2024vyg,Narovlansky:2023lfz,Susskind:2023hnj,Silverstein:2022dfj,Batra:2024kjl}. 
In particular, notable progress has been made in the static patch, where local operators gravitationally dressed to the observer’s worldline offer an explicit realization of relational observables~\cite{Chandrasekaran:2022cip,Witten:2023xze}.  It was shown in~\cite{Chandrasekaran:2022cip} that these dressed operators generate a Type~II$_1$ algebra, denoted $\mathcal{A}_{\text{obs}}$, acting on a maximally entangled state $\Psi_{\text{max}}$. This state consists of the empty de~Sitter vacuum $\Psi_{\text{dS}}$ in the static patch and the thermal equilibrium state of the observer with inverse temperature $\beta_{\text{dS}}$. Furthermore,~\cite{Witten:2023xze} demonstrated that $\Psi_{\text{max}}$ satisfies the KMS condition corresponding to infinite temperature by explicitly analyzing the two-point functions of the dressed operators:
\be
\la\Psi_{\max }|\hat{a} \hat{b}| \Psi_{\max }\ra=\la\Psi_{\max }|\hat{b} \hat{a}| \Psi_{\max }\ra,\quad\forall\hat{a},\hat{b}\in\mathcal{A}_{obs}.
\ee
This is regarded as a distinctive feature of gravitational observables within the static patch. It is also noteworthy that the derivation in~\cite{Witten:2023xze} does not rely on any \emph{a priori} assumption about an infinite-temperature limit.

\justify On the other hand, it has been observed that in the double-scaled SYK
(DSSYK) model, a finite effective temperature emerges in the infinite-temperature
limit, characterizing the thermal behavior of the system in this
regime~\cite{Lin:2022nss,Rahman:2024vyg}. This observation motivates the
proposal of~\cite{Susskind:2023hnj}, which suggests that the infinite-temperature
limit of the DSSYK model describes the confined degrees of freedom residing on
the stretched horizon of de~Sitter space. In this scenario, the bulk physics
emerges at a finite effective temperature from holographic degrees of freedom
that are themselves in an infinite temperature.  A modern review and detailed
analysis of the states and operators in the DSSYK framework have been presented
in~\cite{Lin:2022rbf,Lin_2023}, making extensive use of the chord language.
Building on the chord formulation, several works have further developed the
connection between DSSYK and the representation theory of quantum
groups~\cite{Berkooz:2022mfk,Blommaert:2023opb,Blommaert:2023wad,Almheiri:2024ayc}.
In particular, it has been highlighted that the double-scaled algebra generated
by more than one types of chord operators realizes a Type~II$_1$ von Neumann
algebra, with the empty chord state $\Omega$ representing a maximally entropic
state---closely analogous to the vacuum structure in the static patch of
de~Sitter space. Other interesting limits that lead to different types of algebra
are explored in \cite{Basteiro:2024alg}.

\paragraph{Goal of the Current paper} The current paper aims to put various statements mentioned above on
a firmer ground, by an explicit construction of the double-scaled algebra $\ma=\operatorname{vN}(H_L,M_L)$ in the limiting chord representation, and to prove that it is
indeed a Type~II$_1$ von~Neumann factor, as expected in \cite{Lin:2022rbf} by drawing its connection to the $q$-Gaussian process \cite{Boejko_1997}.\footnote{
The study of observable algebras in JT gravity coupled to matter, which is expected to arise as the triple-scaling limit of the algebra considered here, has been carried out in~\cite{Penington:2023dql,Kolchmeyer:2023gwa}.} In presenting the proof, we declare that the empty state $\Omega$ with no open chords is cyclic and separating for $\ma$, fulfilling the KMS condition of infinite temperature and giving rise to the unique normalized Type~II$_1$ trace $\tau$. This provides a mathematically precise sense in which the empty chord state $|\Omega\rangle$ corresponds to an infinite-temperature state, and it also offers an alternative justification for the following relation:
\be \label{eq:z0}
Z_{0}(\beta) = \langle \Omega | e^{-\beta H_{0}} | \Omega \rangle \stackrel{?}{=} \operatorname{Tr}(e^{-\beta H_{0}}),
\ee
\justify which implicitly assumes that the empty state plays the role of a trace. Notably, the above definition of the partition function was first introduced in~\cite{Berkooz:2018jqr}, where a diagrammatic formulation, often referred to as chord diagrams, was used as a computational tool for evaluating double-scaled quantities. However, this definition is valid only when the empty state is indeed tracial, a property that depends on the structure of the algebra under consideration. As a counterexample, one can readily verify that the empty state fails to be tracial once the chord-number operator is included alongside the chord operators. One of the main goals of the present paper is therefore to establish precisely to what extent the empty state \textbf{does} serve as a trace and how the relation~\eqref{eq:z0} should be interpreted as a partition function.

In conducting the proof, we introduce a convenient notion of normal ordering, which allows us to formulate the cyclic property of $\Omega$ as a bulk-to-boundary correspondence. This correspondence proves especially useful for understanding the bulk dual description of boundary observables and extends the ``complexity equals bulk length'' proposal~\cite{bulkmanifestation2023} to cases with matter chord insertions.

In addition to the algebraic analysis, we provide a complete set of analytic solution for the one-particle wavefunctions, which realize irreducible representations of the symmetry algebra developed in \cite{Lin_2023}. These wavefunctions serve as fundamental building blocks for constructing the full chord Hilbert space. Furthermore, we show how correlation functions can be recovered by taking the inner product with an appropriate matter density, which naturally defines the one-particle inner product structure within the chord Hilbert space.

\paragraph{Organization of the Paper}
In section \ref{sec:pure-DSSYK} we briefly review the construction of Hilbert space $\mh_{0}$
in double-scaled SYK model without matter chords. We present both
the chord number basis and energy basis and the overlap between
them. We discuss two interesting scenarios where in the first case,
the observer has access to all bounded operators $\mathcal{B}\left(\mh_{0}\right)$
and in latter case the observer only has access to bounded functions
of the Hamiltonian $H_{0}$, denoted as $\ma_{0}$. In the first case
the algebra is of Type I$_{\infty}$ and the trace of this algebra
is uniquely defined as summing over expectation values of all basis
states. In the second case the algebra $\ma_{0}$ is not a factor
and there is no preferred definition of a trace. Consequently, in both scenarios, there is no justification for employing the expectation value in $\Omega$ as a trace, as was implicitly assumed in \eqref{eq:z0}.

In section \ref{sec:DSSYK-matter} we construct the double-scaled algebra $\ma$ generated by chord operators
and prove it is a Type II$_{1}$ factor.\footnote{Throughout the paper, we define the double-scaling limit of operators in the renormalized tracial state. It is found recently that for certain pure states, the resulting double-scaled algebra can be a Type I$_\infty$ factor, see \cite{rajgadia2026}. } We specify the
Hilbert space $\mh$ of double-scaled SYK model with a single type
of matter with weight $\Del$, and define operators of $\ma$
by explicitly specifying their action on a generic state in $\mh$.
By construction, $\mh$ contains $\mh_{0}$ as a subspace. We then explore the modular structure of $\ma$ and prove that the empty state $\Omega$ is cyclic separating for $\ma$. Hence, an alternative and equally valid approach would be to initially define the operators' action on $\Omega$ and subsequently applying the GNS construction.  We opt for the current presentation approach as it is directly motivated by the chord statistics established in \cite{Berkooz:2018jqr,Lin:2022nss}. We assert the existence of an operator basis
$ \{\Phi_{\xi}|\thinspace\thinspace\thinspace\forall|\xi\ra\in\mh\}$ of $\ma$ by applying normal orderings to strings of chord operators.
Subsequently, we formulate the cyclic property of $\Omega$ in terms of the operator-state correspondence as follows:
\be
\Phi_{\xi}|\Omega\ra=|\xi\ra.
\ee
We will utilize this operator language when addressing the finite emergent temperature in the following sections.

In section \ref{sec:limits} we explore various limits of double-scaled SYK model
and relations to JT gravity, theory of baby universe, and Brownian
double-scaled SYK. In section \ref{sec:triple} we revisit the triple scaling limit
explored in \cite{Lin:2022rbf} and extend the discussion of resulting Liouville quantum mechanics to one-particle sector. Various
relations among gauge invariant wavefunctions in JT gravity can be
derived by taking triple scaling limit of their counterpart in DSSYK.
We present two solvable scenarios for the one-particle wavefunction and provide insights on the expectations in a more general situation. In section \ref{sec:baby}
we consider the $q\to1$ limit of DSSYK with the penalty factor $r$ for $M$-$M$ crossing and $r_{V}$ for $H$-$M$ crossing fixed
as independent parameters \footnote{This setup extends the conventional parameter regime of the double-scaled SYK (DSSYK) model, where the penalty factors $r$ and $r_V$ are related to the deformation parameter $q$ through the matter scaling dimension. More precisely, the relations are given by $r_V = q^{\Delta_V}, \quad r = q^{\Delta_V^2}.$
The chord Hilbert space can be consistently constructed by assuming that the parameters $q$, $r_V$, and $r$ are real numbers lying within the interval $[0,1)$.}. We present an explicit expression of inner
product between states with arbitrary amount of matter and Hamiltonian
chords. In a specific instance, we illustrate how chord dynamics closely resembles the behavior seen in the theory of baby universes in the semi-classical limit. This involves processes such as splitting and rejoining, or the direct evolution of baby universes from their initial to final states \cite{Giddings:1988cx}. We conclude the section with an alternative
presentation of the inner product, where the sum-over-matrices involved
are achieved with help of an integral implementation of constraints.
In section \ref{sec:BDSSYK} we comment on the relation between $q\to0$ limit with
$r$ and $r_{V}$ fixed and Brownian DSSYK developed in \cite{Milekhin:2023bjv}. We present the expression of the inner product in this limit
and explore the corresponding relations to states and algebra in Brownian DSSYK. 

In section \ref{sec:future} we discuss various future directions of the algebraic
study of DSSYK. In particular, we reformulate the results of the inner product by expressing them in terms of correlation functions of operators within $\ma$ under the triple scaling limit.  Note that it is a correlation function
in an infinite temperature state $|\Omega\ra$, but the result exhibits explicit dependence on a finite temperature parameter $c=c(\beta)$. We intend to view this fact as a preliminary manifestation of the idea that a finite effective temperature emerges, serving to characterize the thermal behavior of the system within an infinite-temperature state. The dependence of this effective temperature is encoded in the operator algebra, despite the fact that the state $\Omega$ exhibits infinite temperature.  We leave future exploration of this point and a potential algebraic characterization of hyper-fast scrambling to future work. 

We present various details that are used in the main text in appendices. In appendix~\ref{app:review}, we review the basic concepts of von~Neumann factors and their classification.  In appendix \ref{app:full-solution}, we present a full solution of energy spectrum in DSSYK
with a generating function method. In particular, the one-particle
irreducible representations can be solved in terms of eigenstates
of the left and right Hamiltonian. We further show that the inner
product between $1$-particle states can be reproduced by inserting
an energy eigenbasis, and integrating the left and right energy with
proper measure. In appendix \ref{app:Fock} we comment on the relations among Lin-Stanford chord 
basis and Fock basis. We present an alternative formulation of $1$-particle
inner product in terms of matter correlators in the Fock basis. In appendix \ref{app:Bound-PQ} we provide derivation of an operator bound on the $Q$-symmetrizer.  In
appendix \ref{app:derivation} we present a detailed derivation of left and right Liouville
Hamiltonian with matter emerging from triple scaling limit. 

\paragraph{Note added in v7.}
The proof of factoriality in an earlier version of this paper used a formal expansion of a putative central operator in the normal-ordered chord basis, together with an \(L^2\)-summability assumption on the corresponding coefficients. This coefficientwise argument captures the correct obstruction: commuting a nontrivial chord component with the generators produces further chord components that cannot be consistently cancelled. However, the argument did not fully justify passing from this formal \(L^2\) expansion to an arbitrary element of the von~Neumann algebra.

The revised proof in section~\ref{sec:proof} fills this gap by using the exponential decay of mixed two-sided  ladder commutators on high chord-number sectors. This gives a quantitative tail estimate showing that an infinite chord-number tail cannot evade the coefficientwise obstruction. Consequently any central vector must lie in the intersection of the pure Hamiltonian- and pure matter-chord subspaces, which is only the vacuum, and the center is trivial.

\paragraph{Glossary}
\begin{itemize}
\item $q\equiv e^{-\lambda},\lambda>0$ gives the penalty factor for crossing Hamiltonian
chords.
\item $r_{V}$ gives the penalty factor for crossing between a Hamiltonian
chord and a matter chord. We keep it independent of $q$ in most context
of the current paper and specify $r_{V}=q^{\Del_{V}}$ only in certain
context. 
\item $r$ gives the penalty factor for crossing matter chords. We keep
it independent of $q$ in most context of the current paper and specify
$r=q^{\Del_{V}^{2}}$ only in certain context. 
\item $[A,B]_{q}\equiv AB-qBA$  is the $q$-commutator.
\item $[n]_{q}\equiv\frac{1-q^{n}}{1-q}=1+q+q^2+\dots +q^{n-1}$ is the $q$-integer. 
\item $\left(a;q\right)_{n}\equiv\prod_{k=1}^{n}\left(1-aq^{k-1}\right)$ is
the $q$-Pochhammer symbol. $(q;q)_0 \equiv 1$.
\item $\left(a_{1},a_{2}\cdots,a_{k};q\right)_{\infty}=\prod_{i=1}^{k}\left(a_{i};q\right)_{\infty}$
\item $[n]_{q}!=\left(q;q\right)_{n}/\left(1-q\right)^{n}$ is the $q$-factorial
\item $\mh_{0}$: The Hilbert space of DSSYK without matter chords.
\item $\mh$: The Hilbert space of DSSYK that contains matter chords. In
the current paper we only consider a single type of matter, characterized
by $r_{V}$ and $r$. 
\item $|\Omega\ra$ or $\Omega$: Empty state with no open chords. 
\item $|n_{0},n_{1},\cdots,n_{k}\ra$: A typical state in $\mathcal{H}$ consists of $k$ matter chords, with $n_{i-1}$ Hamiltonian chords positioned between the $\left(i-1\right)$-th and $i$-th matter chord, where $i$ ranges from 1 to $k$.
\item $a_{L}^{\da}$: creates a Hamiltonian chord from the left: $a_{L}^{\da}|n_{0},\cdots,n_{k}\ra=|n_{0}+1,\cdots,n_{k}\ra$. 
\item $a_{L}$: defined as Hermitian conjugate of $a_{L}^{\da}$. Its action on state is defined in \eqref{eq:ladder-1}.
\item $b_{L}^{\da}$: creates a matter chord from the left: $b_{L}^{\da}|n_{0},\cdots,n_{k}\ra=|0,n_{0},\cdots,n_{k}\ra$. 
\item $b_{L}$: defined as Hermitian conjugate of $b_{L}^{\da}$. Its action on state is defined in \eqref{eq:ladder-2}.
\item $H_{L}\equiv a_{L}+a_{L}^{\da}$ is the left Hamiltonian chord operator. 
\item $M_{L}\equiv b_{L}+b_{L}^{\da}$ is the left matter chord operator. 
\item $\ma_{L}:$ The left double-scaled algebra generated by completion of finite linear span of strings with two letters: $H^{n_0}_L M_L H^{n_1}_L\cdots M_L H^{n_k}_L$.
\item $\Phi_L \left(n_{0},\cdots,n_{k}\right)$: The left chord field operator that satisfies
$\Phi_L\left(n_{0},\cdots,n_{k}\right)|\Omega\ra=|n_{0},\cdots,n_{k}\ra$.
\item $H_{n}\left(x|q\right)$: $q$-Hermite polynomial of order $n$. 
\item $H_{m,n}\left(x,y|q,r_V \right)$ bivariate $r_V$-weighted $q$-Hermite
polynomial of order $\left(m,n\right)$.
\item $K_{\nu}\left(x\right)$: modified Bessel function of the second kind with order
$\nu$. 
\item $\mathcal{B}(\mathcal{H})$: The Von Neumann algebra of bounded linear operators acting on Hilbert space $\mathcal{H}$.
\item $|\te\ra$: energy eigenbasis in $\mh_{0}$.
\item $\mu\left(\te\right)\equiv\left(2\pi\right)^{-1}\left(e^{\pm2i\te},q;q\right)_{\infty}$
is the measure that defines inner product between energy basis. $\la\te_{1}|\te_{2}\ra=\mu\left(\te_{1}\right)^{-1}\del\left(\te_{1}-\te_{2}\right)$.
\end{itemize}

\section{Warm up: States and Algebra in DSSYK without Matter Chords} \label{sec:pure-DSSYK}
In this section, we summarize the essential ingredients of the states and operators in the DSSYK model without matter.  
For a modern review of this subject, particularly the details of how the double-scaling limit leads to a chord description of SYK operators, we refer the reader to~\cite{Berkooz:2024review}.The Hilbert space is spanned by states containing a definite number of Hamiltonian chords.   For a state with a fixed chord number, the Hamiltonian, often referred to as the $H$-chord operator\footnote{
In the language of \cite{Susskind:2023hnj}, Hamiltonian chords are named as chords while matter chords, which we introduce in subsequent section,}  are referred to as chords. It acts by either creating a new chord or removing an existing one, thereby forming a closed chord.   Importantly, the annihilation of chords depends sensitively on the position of the chord being removed, as one must take into account the number of crossings being created in the process.   This mechanism leads to the $q$-deformed commutation relations between the chord creation and annihilation operators, providing an operator formulation of the chord statistics developed in~\cite{Berkooz:2018jqr}, which we elaborate on in this section.

The $H$-chord operator is defined in terms of  the $q$-ladder operators:
\be\label{eq:H0-def}
H_{0}=a+a^{\da}.
\ee
where $a$ and $a^{\da}$ satisfy the $q$-commutator:
\be
[a,a^{\da}]_{q}=aa^{\da}-qa^{\da}a=1.
\ee
We used $H_0$ to refer to the $H$-chord operator that acts on the Hilbert space without matter.  The spectrum of the theory is specified by states with different number
of chords:
\be
a^{\da}|n\ra=|n+1\ra,\quad a|n\ra=[n]|n-1\ra,\qq a|0\ra=0.
\ee
where $[n]$ is a $q$-deformed integer defined as follows:
\be
[n]=q[n-1]+1,\qq[n]=\frac{1-q^{n}}{1-q},\qq[n]!=\frac{\left(q;q\right)_{n}}{\left(1-q\right)^{n}},
\ee
where $[n]!=\prod_{k=1}^{n} [k]$ is the $n$-factorial. it is then straightforward to show 
\be
\la n|m\ra=\del_{nm}[n]!.
\ee
These states are unnormalized.  When an orthonormal number basis is useful, we write
\be \label{eq:normalized-number-basis}
|n\ra_{N}:=\frac{|n\ra}{\sqrt{[n]!}},
\qquad
{}_{N}\!\la n|m\ra_{N}=\del_{nm}.
\ee
We will use $\varphi_n(\te)=\la\te|n\ra$ for the overlap with the unnormalized state and $\psi_n(\te)=\la\te|n\ra_N$ for the normalized wavefunction.
In chord language, the preparation of state $|n\ra$ is achieved by slicing open a chord diagram with $n$ non-intersecting open Hamiltonian chords, see \eqref{eq:example-3} for an illustration.
The inner product between a bra state $\la m|$ and a ket state $|n\ra$
is defined by sewing the two open diagrams into a joint one, and sum
over all possible pairings of open chords between the initial and final
state. For each given pairing, the result is weighted by $q^{c}$
where $c$ is the amount of crossings in the chord configuration determined
by the pairing. Therefore, if $m\not=n$,
one cannot pair all open chords, leading to vanishing result. When
$n=m$, the result is $[n]!$ which correctly counts the $q$-weighted
sum.

An alternative way of understanding the origin of $[n]!$ is to think
of the $q$-weighted sum above as a summation over element $\pi$
in permutation group $S_{n}$ with a discrete measure $q^{\iota\left(\pi\right)}$,
where $\iota\left(\pi\right)$ counts the inversions in $\pi$. A
given configuration of chords with fixed initial and final states can be mapped to an element $\pi$
in $S_{n}$, and the amount of pairings is counted by the inversions
in $\pi$. As an example, let us consider the case $n=3$:
\begin{equation}\label{eq:example-3}
\begin{tikzpicture}[baseline={([yshift=-0.1cm]current bounding box.center)},scale=1]
     \draw[thick] (0,0) arc (0:180: 2);
    \draw[thick] (0,-0.5) arc (0:-180: 2);
    \draw[thick,blue,dashed] (-4.5,0) -- (0.5,0);
    \draw[thick,blue,dashed] (-4.5,-0.5) -- (0.5,-0.5);
    \draw (-1,0) -- (-1,1.73);
    \draw (-2,0) -- (-2,2);
    \draw (-3,0) -- (-3,1.73);
    \draw (-1,-0.5) -- (-1,-2.23);
    \draw (-2,-0.5) -- (-2,-2.5);
    \draw (-3,-0.5) -- (-3,-2.23);
    \node at (-5,1) {$\la 3 | = $};
    \node at (-5,-1.5) {$|3\ra =$};
    \node at (-1,1.73) [circle,fill,black,inner sep=1.2pt]{};    
    \node at (-2,2) [circle,fill,black,inner sep=1.2pt]{};
    \node at (-3,1.73) [circle,fill,black,inner sep=1.2pt]{};    
    \node at (-1,-2.23) [circle,fill,black,inner sep=1.2pt]{};    
    \node at (-2,-2.5) [circle,fill,black,inner sep=1.2pt]{};
    \node at (-3,-2.23) [circle,fill,black,inner sep=1.2pt]{};
\end{tikzpicture}=
\sum_{\pi\in S_3} \begin{tikzpicture}[baseline={([yshift=-0.1cm]current bounding box.center)},scale=1]
    \draw[thick] (0,0) arc (0:180: 2);
    \draw[thick] (0,-0.5) arc (0:-180: 2);
    \draw[thick,blue,dashed] (-4.5,0) -- (0.5,0);
    \draw[thick,blue,dashed] (-4.5,-0.5) -- (0.5,-0.5);
    \draw (-1,0) -- (-1,1.73);
    \draw (-2,0) -- (-2,2);
    \draw (-3,0) -- (-3,1.73);
    \draw (-1,-0.5) -- (-1,-2.23);
    \draw (-2,-0.5) -- (-2,-2.5);
    \draw (-3,-0.5) -- (-3,-2.23);
    \node at (-0.8,1.95) {$\pi(1)$};
    \node at (-2,2.3) {$\pi(3)$};
    \node at (-3.2, 1.95) {$\pi(2)$};
    \node at (-0.8, -2.45) {$3$};
    \node at (-2,-2.8) {$2$};
    \node at (-3.2, -2.45) {$1$};
    \draw[red] (-1,0)--(-3,-0.5);
    \draw[red] (-2,0)--(-1,-0.5);
    \draw[red] (-3,0)--(-2,-0.5);    
    \node at (-1,1.73) [circle,fill,black,inner sep=1.2pt]{};    
    \node at (-2,2) [circle,fill,black,inner sep=1.2pt]{};
    \node at (-3,1.73) [circle,fill,black,inner sep=1.2pt]{};    
    \node at (-1,-2.23) [circle,fill,black,inner sep=1.2pt]{};    
    \node at (-2,-2.5) [circle,fill,black,inner sep=1.2pt]{};
    \node at (-3,-2.23) [circle,fill,black,inner sep=1.2pt]{};
\end{tikzpicture}.    
\end{equation}
There are six ways of pairing
the chords in initial and final state, each corresponds to an element $\pi$ in $S_3$. Different $\pi$ corresponds to a $3$-permutation, and we can specify it by its image: $(\pi(1),\pi(2),\pi(3))\in \{(123),(231),(312),(213),(132),(321) \}=S_3$, with corresponding amount of inversions $\{0,2,2,1,1,3\}$. Therefore we find in this case:
\be
\sum_{\pi\in S_3} q^{\iota(\pi)}=(1+q)(1+q+q^2)=[3]_q !,
\ee
which reproduces the sum over intermediate crossings in \eqref{eq:example-3}.
More generally, we have
\be
\sum_{\pi\in S_{n}}q^{\iota\left(\pi\right)}=[n]!.
\ee
We shall adopt this formulation of inner product in the discussion
involving matter. In conclusion, one can define the Hilbert space of DSSYK
without matter as
\be
\mh_{0}=\oplus_{n=0}^{\infty}\mathbb{C}|n\ra.
\ee
with inner product specified above. 

it is also helpful to introduce
energy basis $|\te\ra$. The energy basis was originally found in
\cite{Berkooz:2018qkz} by diagonalizing $H_{0}$ with an infinite transfer matrix,
and correctly evaluate the normalization factor. We briefly summarize
the result as follows. The action of $H_{0}$ on $|\te\ra$ is given
by:
\be
H_{0}|\te\ra=E_{0}\left(\te\right)|\te\ra=\frac{2\cos\te}{\sqrt{1-q}}|\te\ra,\qq\te\in\left[0,\pi\right].
\ee
Clearly the energy spectrum is compactly supported in $[-\frac{2}{\sqrt{1-q}},\frac{2}{\sqrt{1-q}}]$.
The inner product between energy eigenstate is:
\be
\la\te_{1}|\te_{2}\ra=\frac{\del\left(\te_{1}-\te_{2}\right)}{\mu\left(\te_{1}\right)},\quad\mu\left(\te\right)=\frac{\left(e^{\pm2i\te},q;q\right)_{\infty}}{2\pi}.
\ee
The overlaps of the energy basis with the unnormalized and normalized chord-number states are
\be \label{eq:number-energy-overlaps}
\varphi_n(\te):=\la\te|n\ra
=\frac{H_n(\cos\te|q)}{(1-q)^{n/2}},
\qquad
\psi_n(\te):=\la\te|n\ra_N
=\frac{H_n(\cos\te|q)}{\sqrt{(q;q)_n}},
\ee
with $\varphi_0=\psi_0=1$.  Accordingly,
\be
\int_0^\pi \mu(\te)\dd\te\,\varphi_n(\te)\varphi_m(\te)
=\del_{nm}[n]!,
\qquad
\int_0^\pi \mu(\te)\dd\te\,\psi_n(\te)\psi_m(\te)
=\del_{nm}.
\ee
As a direct application, The partition function\footnote{The rationale behind defining $Z_0 (\beta)$ as the expectation value of $e^{-\beta H_0}$ in the state $|0\ra$ remains unclear at the moment.  We comment on this question at the end of this section and address it in subsequent sections. } can be evaluated as
\be \label{eq:partition}
\begin{split}
Z_0\left(\beta\right) & =\la0|e^{-\beta H}|0\ra=\int_{0}^{\pi}\mu\left(\te\right)|\psi_{0}\left(\te\right)|^{2}\dd\te e^{-\frac{2\beta\cos\te}{\sqrt{1-q}}}\\
 & =\frac{\sqrt{1-q}}{\beta}\sum_{\nu=0}^{\infty}(-1)^{\nu}q^{\frac{\nu\left(\nu+1\right)}{2}}(2\nu+1)I_{2\nu+1}\left(\frac{2\beta}{\sqrt{1-q}}\right),
\end{split}
\ee
where $I_{\nu}\left(x\right)$ is the modified Bessel function of
the first kind. The result is valid for arbitrary inverse temperature $\beta$
and $q: |q|<1$. 

In terms of energy basis, we can define $\mh_{0}$ alternatively as
all $L^{2}$-integrable functions in $[0,\pi]$ with measure $\mu\left(\te\right)$:
\be
\mh_{0}=L^{2}\left([0,\pi],\mu(\te)\right).
\ee
Now we move on to the discussion of the operator algebra. One situation is that
the observer has full access to all bounded operators that act on
$\mh_{0}$, namely, in this case $\ma_{obs}=\mathcal{B}(\mh_{0})$.
This is a Type~I$_\infty$ von~Neumann algebra, namely the algebra of bounded operators on an infinite-dimensional separable Hilbert space.  It carries the canonical faithful normal semifinite Hilbert-space trace.  For a positive or trace-class operator $a\in\mathcal{B}(\mh_0)$,
\be \label{eq:trace1}
\operatorname{Tr}_{\mh_0}(a)
:=\sum_{n=0}^{\infty}\frac{\la n|a|n\ra}{[n]!}
=\sum_{n=0}^{\infty}{}_{N}\!\la n|a|n\ra_N
=\int_{0}^{\pi}\mu\left(\te\right)\la\te|a|\te\ra\dd\te,
\ee
whenever the diagonal kernel in the last expression is defined.
In particular, the observer can measure the amount of chords in state
$|n\ra$ by looking at the expectation value of the size operator
$q^{\hat{n}}$. This is a bounded operator with discrete spectrum
in $[0,1]$, and is a trace-class operator with
\be
\operatorname{Tr}_{\mh_0}\left(q^{\hat{n}}\right)=\sum_{n=0}^{\infty}q^{n}=\frac{1}{1-q}.
\ee
There is another interesting situation where the observer has only
access to $H_{0}$, or more concretely, all operators that are functions
of $H_{0}$. In this case, the observer algebra $\ma_{obs}$ is the
maximal commutative Von Neumann subalgebra of $\mathcal{B}\left(\mathcal{H}\right)$
that contains $H_{0}$, which we denote as $\ma_{0}$. 

With operators in $H_{0}$, the observer can access the state with arbitrary chords 
by adding chords into the empty state $|0\ra$:
\be
|1\ra=H_{0}|0\ra,\qq|2\ra=H_{0}^{2}|0\ra-|0\ra,\dots,
\ee
however, in this case the observer would not be able to know the trace
defined in \eqref{eq:trace1}. This is because for a commutative algebra such
as $\ma_{0}$, any faithful positive linear functional $p:\ma_{0}\to\mathbb{C}$
satisfies
\be
p\left(ab\right)=p\left(ba\right),\qq\forall a,b\in\ma_{0}.
\ee
As a result, they are equally valid to the observer as a trace. This is similar to the situation in pure JT gravity where the only gauge invariant boundary operators are functions of the left or right Hamiltonian \cite{Penington:2023dql}. They are forced to be equivalent $H^{JT}_L=H^{JT}_R=H^{JT}_0$ after gauging the $SL(2,\mathbb{R})$ symmetry. Consequently, the theory lacks a preferred choice of trace unless an additional independent assumption is incorporated into its definition. $\mathcal{A}_0$ is not a von~Neumann factor, as its center is nontrivial and, owing to its commutative nature, coincides with the algebra itself.

In either situation above, there is no natural reason to define the
trace of the theory to be the expectation value in the empty state,
as we did in defining the partition function $Z_0\left(\beta\right)$
as in \eqref{eq:partition}. In the following section, we will enlarge $\ma_{obs}$ by incorporating matter chord operators. As a result, the empty state characterized by the absence of both Hamiltonian chords and matter chords, becomes the tracial state unique up to constant rescaling for the extended algebra. Consequently, there is a preferred normalized trace $\tau$ on the extended algebra, and one can reformulate \eqref{eq:partition} as 
\be
Z_0 (\beta)= \la\Omega |e^{-\beta H_0 }|\Omega\ra = \tau(e^{-\beta H_0}). 
\ee

\section{States and Algebra in DSSYK with Matter Chords}  \label{sec:DSSYK-matter}
In previous section we established Hilbert space description of the
dynamics of Hamiltonian chords. In this section we introduce operators
that generate matter chords in the state. For simplicity, we only
consider one species of matter chord in the following discussion,
and the analysis for multiple species follows in a similar manner.
\subsection{Construction of the Double-Scaled Algebra}\label{sec:construction}
We construct the Hilbert space $\mh$ by tensoring with the space of chord states that contains multiple particles, which is similar\footnote{This is not the standard Fock space construction because the states are not defined as simple tensor product of those in $\mh_0$. To distinguish, we call the basis $|n_0,\cdots,n_k\ra $ in $\mh$ the Lin-Stanford basis and refer the reader to appendix \ref{app:Fock} for exploration of its relation to the Fock basis $|n_0\ra\otimes\cdots\otimes |n_k\ra$. } to a Fock space construction of $\mh_0$:
\be
\mh=\overline{\bigoplus_{k=0}^{\infty}\operatorname{span}_{\mathbb{C}}
\left\{|n_{0},\dots,n_{k}\ra:(n_{0},\dots,n_{k})\in\mathbb{N}^{k+1}\right\}},
\ee
In literature, the state $|0\ra$ is alternatively referred to as $|\Omega\ra$ emphasizing its role as the empty state. In the following discussion, we will consistently employ the notation $\Omega$ to represent this state. A general state in $\mh$ can then be denoted as $|n_{0},\cdots,n_{k}\ra$, which corresponds to a ket state with $k$ open matter chords, which separates the half-chord diagram into $k+1$ divisions, and there are $n_i$ Hamiltonian chords in the $i$-th division. 
The inner product for $\mh$ is defined as
\be \label{eq:inner-product}
\la n_{0},n_{1},\cdots,n_{k}|m_{0},m_{1},\cdots,m_{l}\ra=\del_{kl}\sum_{\pi\in S_{k}}r^{\iota \left(\pi\right)}\la n_{0},n_{1},\cdots,n_{k}|m_{0},m_{1},\cdots,m_{l}\ra^{\pi},
\ee
where $\iota\left(\pi\right)$ is the number of inversion in permutation
$\pi$, and $r$ is the penalty factor for a crossing between two matter chords. In the following
discussion, we keep $r$ as a independent parameter of $q$ and ranges
from $r\in(0,1)$. The reason for summing over the permutations $\pi$
in $S_{k}$ in the inner product \eqref{eq:inner-product} is to include general configurations of matter chord intersections in the inner product. The permutation dependent inner product is
defined recursively as:
\be \label{eq:inner-product2}
\la  n_{0},\cdots,n_{k}|m_{0},\cdots,m_{l}\ra^{\pi}=\sum_{j=0}^{l}[m_{j}]q^{\sum_{j^{\pp}<j}m_{j}}r_{V}^{j}\la n_{0}-1,\cdots,n_{k}|m_{0},\cdots m_{j}-1,\cdots,m_{k}\ra^{\pi},
\ee
with boundary condition: 
\be\begin{split} \label{eq:inner-product-bc}
\la0,\cdots,n_{i},\cdots,0|0,\cdots,n_{i},\cdots,0\ra^{\pi} & =[n_{i}]_{q}!r_{V}^{2c_{\pi}\left(i\right)n_{i}},\qq i=0,1,\cdots k,
\end{split}\ee
where the function of permutations $c_{\pi}\left(i\right)$ is defined as 
\be
c_{\pi}\left(i\right)=\#\{\pi(j)|j\leq i,\qq i<\pi\left(j\right)\},
\ee
which counts the extra crossings between Hamiltonian chords and matter chords in a given channel. To illustrate, let us consider the case $\pi:\left(1,2,3,4\right)\to\left(4,3,2,1\right)$,
we know $c_{\pi}\left(1\right)=1,c_{\pi}\left(2\right)=2,c_{\pi}\left(3\right)=1$
and $c_{\pi}\left(0\right)=c_{\pi}\left(4\right)=0$. $r_{V}$ in
\eqref{eq:inner-product-bc} is the penalty factor for a single matter-Hamiltonian crossing, with its exponent counting the amount of such crossings. Similar to $r$, we treat $r_{V}$ as an independent parameter that ranges from $\left(0,1\right)$ in the following discussion.

 The evaluation of \eqref{eq:inner-product2} can be understood as follows: For a fixed permutation $\pi\in S_k$, we have a chord diagram with a matter chord background corresponding to $\pi$, see \eqref{eq:example-4} for an illustration. The amount of crossings among matter chords is counted by inversions of $\pi$. Then we insert Hamiltonian chords into this background, and prepare the bra and ket states by specifying the amount of Hamiltonian chords in between each matter chords. The inner product \eqref{eq:inner-product}  then sums over all Hamiltonian chords configuration in a given matter chord background determined by $\pi$, and then sums over all permutations $\pi\in S_n$.  The case with $3$-particles can be visualized as: 
\begin{equation}\label{eq:example-4}
\begin{tikzpicture}[baseline={([yshift=-0.1cm]current bounding box.center)},scale=0.8]
     \draw[thick] (0,0) arc (0:180: 2);
    \draw[thick] (0,-0.5) arc (0:-180: 2);
    \draw[thick,blue,dashed] (-4.5,0) -- (0.5,0);
    \draw[thick,blue,dashed] (-4.5,-0.5) -- (0.5,-0.5);
    \draw[thick,blue] (-1,0) -- (-1,1.73);
    \draw[thick,blue] (-2,0) -- (-2,2);
    \draw[thick,blue] (-3,0) -- (-3,1.73);
    \draw[thick,blue] (-1,-0.5) -- (-1,-2.23);
    \draw[thick,blue] (-2,-0.5) -- (-2,-2.5);
    \draw[thick,blue] (-3,-0.5) -- (-3,-2.23);
    \node at (-6.3,1) {$\la n_0,n_1,n_2,n_3 | = $};
    \node at (-6,-1.5) {$|m_0,m_1,m_2,m_3\ra =$};
    \node at (-1,1.73) [circle,fill,blue,inner sep=1.2pt]{};    
    \node at (-2,2) [circle,fill,blue,inner sep=1.2pt]{};
    \node at (-3,1.73) [circle,fill,blue,inner sep=1.2pt]{};    
    \node at (-1,-2.23) [circle,fill,blue,inner sep=1.2pt]{};    
    \node at (-2,-2.5) [circle,fill,blue,inner sep=1.2pt]{};
    \node at (-3,-2.23) [circle,fill,blue,inner sep=1.2pt]{};
    \draw (-3.92,0) -- (-3.92,0.544);
    \draw (-3.71,0)--(-3.71,1.033);
    \draw (-3.27,0)--(-3.27,1.538);
    \draw (-3.497,0)--(-3.497,1.326);
    \draw (-2.852,0)--(-2.852,1.81);
    \draw (-2.59,0)--(-2.59,1.909);
    \draw (-2.852,0)--(-2.852,1.81);
    \draw (-1.712,0)--(-1.712,1.979);  
    \draw (-1.89,0)--(-1.89,1.997);
    \draw (-1.42,0)--(-1.42,1.91);
    \draw (-0.987,0)--(-0.987,1.67);
    \draw (-0.9,0)--(-0.9,1.67);
    \draw (-0.485,0)--(-0.485,1.306);
    \node at (-3.92,0.544) [circle,fill,black,inner sep=1pt]{};
    \node at (-3.71,1.033) [circle,fill,black,inner sep=1pt]{};
    \node at (-3.27,1.538) [circle,fill,black,inner sep=1pt]{};
    \node at (-3.497,1.326) [circle,fill,black,inner sep=1pt]{};
    \node at (-2.852,1.81) [circle,fill,black,inner sep=1pt]{};
    \node at (-2.59,1.909) [circle,fill,black,inner sep=1pt]{};
    \node at (-1.712,1.979) [circle,fill,black,inner sep=1pt]{};
    \node at (-1.89,1.997) [circle,fill,black,inner sep=1pt]{};
    \node at (-1.42,1.91) [circle,fill,black,inner sep=1pt]{};
    \node at (-0.987,1.67) [circle,fill,black,inner sep=1pt]{};
    \node at (-0.9,1.67) [circle,fill,black,inner sep=1pt]{};
    \node at (-0.485,1.306) [circle,fill,black,inner sep=1pt]{};
    \draw (-3.78,-0.5) -- (-3.78,-1.404);
    \draw (-3.49,-0.5)--(-3.49,-1.827);
    \draw (-3.14,-.5)--(-3.14,-2.14);
    \draw (-2.75,-0.5) -- (-2.75,-2.35);
    \draw (-2.29,-0.5)--(-2.29,-2.478);
    \draw (-2.47,-.5) -- (-2.47,-2.44);
    \draw (-1.26,-.5) -- (-1.26,-2.36);
    \draw (-1.558,-0.5)--(-1.558,-2.45);
    \draw (-.809,-.5) -- (-.809,-2.106);
    \draw (-.69,-.5) -- (-.69,-2.017);
    \draw (-.51,-0.5)--(-.51,-1.838);
    \node at (-3.78,-1.404) [circle,fill,black,inner sep=1pt]{};
    \node at (-3.49,-1.827) [circle,fill,black,inner sep=1pt]{};
    \node at (-3.14,-2.14) [circle,fill,black,inner sep=1pt]{};
    \node at (-2.75,-2.35) [circle,fill,black,inner sep=1pt]{};
    \node at (-2.47,-2.44) [circle,fill,black,inner sep=1pt]{};
    \node at (-2.29,-2.478)[circle,fill,black,inner sep=1pt]{};
    \node at (-1.26,-2.36) [circle,fill,black,inner sep=1pt]{};
    \node at (-1.558,-2.45) [circle,fill,black,inner sep=1pt]{};
    \node at (-.809,-2.106) [circle,fill,black,inner sep=1pt]{};
    \node at (-.69,-2.017) [circle,fill,black,inner sep=1pt]{};
    \node at (-.51,-1.838) [circle,fill,black,inner sep=1pt]{};  
\end{tikzpicture}=
\sum_{\pi\in S_3} \sum_{\text{configurations}}\begin{tikzpicture}[baseline={([yshift=-0.1cm]current bounding box.center)},scale=0.8]
    \draw[thick] (0,0) arc (0:180: 2);
    \draw[thick] (0,-0.5) arc (0:-180: 2);
    \draw[thick,blue,dashed] (-4.5,0) -- (0.5,0);
    \draw[thick,blue,dashed] (-4.5,-0.5) -- (0.5,-0.5);
    \draw[thick,blue] (-1,0) -- (-1,1.73);
    \draw[thick,blue] (-2,0) -- (-2,2);
    \draw[thick,blue] (-3,0) -- (-3,1.73);
    \draw[thick,blue] (-1,-0.5) -- (-1,-2.23);
    \draw[thick,blue] (-2,-0.5) -- (-2,-2.5);
    \draw[thick,blue] (-3,-0.5) -- (-3,-2.23);
    \node at (-0.8,1.95) {$\pi(1)$};
    \node at (-2,2.3) {$\pi(3)$};
    \node at (-3.2, 1.95) {$\pi(2)$};
    \node at (-0.8, -2.45) {$3$};
    \node at (-2,-2.8) {$2$};
    \node at (-3.2, -2.45) {$1$};
    \draw[thick,blue] (-1,0)--(-3,-0.5);
    \draw[thick,blue] (-2,0)--(-1,-0.5);
    \draw[thick,blue] (-3,0)--(-2,-0.5);    
    \node at (-1,1.73) [circle,fill,blue,inner sep=1.2pt]{};    
    \node at (-2,2) [circle,fill,blue,inner sep=1.2pt]{};
    \node at (-3,1.73) [circle,fill,blue,inner sep=1.2pt]{};    
    \node at (-1,-2.23) [circle,fill,blue,inner sep=1.2pt]{};    
    \node at (-2,-2.5) [circle,fill,blue,inner sep=1.2pt]{};
    \node at (-3,-2.23) [circle,fill,blue,inner sep=1.2pt]{};
    \draw (-3.92,0) -- (-3.92,0.544);
    \draw (-3.71,0)--(-3.71,1.033);
    \draw (-3.27,0)--(-3.27,1.538);
    \draw (-3.497,0)--(-3.497,1.326);
    \draw (-2.852,0)--(-2.852,1.81);
    \draw (-2.59,0)--(-2.59,1.909);
    \draw (-2.852,0)--(-2.852,1.81);
    \draw (-1.712,0)--(-1.712,1.979);  
    \draw (-1.89,0)--(-1.89,1.997);
    \draw (-1.42,0)--(-1.42,1.91);
    \draw (-0.987,0)--(-0.987,1.67);
    \draw (-0.9,0)--(-0.9,1.67);
    \draw (-0.485,0)--(-0.485,1.306);
    \node at (-3.92,0.544) [circle,fill,black,inner sep=1pt]{};
    \node at (-3.71,1.033) [circle,fill,black,inner sep=1pt]{};
    \node at (-3.27,1.538) [circle,fill,black,inner sep=1pt]{};
    \node at (-3.497,1.326) [circle,fill,black,inner sep=1pt]{};
    \node at (-2.852,1.81) [circle,fill,black,inner sep=1pt]{};
    \node at (-2.59,1.909) [circle,fill,black,inner sep=1pt]{};
    \node at (-1.712,1.979) [circle,fill,black,inner sep=1pt]{};
    \node at (-1.89,1.997) [circle,fill,black,inner sep=1pt]{};
    \node at (-1.42,1.91) [circle,fill,black,inner sep=1pt]{};
    \node at (-0.987,1.67) [circle,fill,black,inner sep=1pt]{};
    \node at (-0.9,1.67) [circle,fill,black,inner sep=1pt]{};
    \node at (-0.485,1.306) [circle,fill,black,inner sep=1pt]{};
    \draw (-3.78,-0.5) -- (-3.78,-1.404);
    \draw (-3.49,-0.5)--(-3.49,-1.827);
    \draw (-3.14,-.5)--(-3.14,-2.14);
    \draw (-2.75,-0.5) -- (-2.75,-2.35);
    \draw (-2.29,-0.5)--(-2.29,-2.478);
    \draw (-2.47,-.5) -- (-2.47,-2.44);
    \draw (-1.26,-.5) -- (-1.26,-2.36);
    \draw (-1.558,-0.5)--(-1.558,-2.45);
    \draw (-.809,-.5) -- (-.809,-2.106);
    \draw (-.69,-.5) -- (-.69,-2.017);
    \draw (-.51,-0.5)--(-.51,-1.838);
    \node at (-3.78,-1.404) [circle,fill,black,inner sep=1pt]{};
    \node at (-3.49,-1.827) [circle,fill,black,inner sep=1pt]{};
    \node at (-3.14,-2.14) [circle,fill,black,inner sep=1pt]{};
    \node at (-2.75,-2.35) [circle,fill,black,inner sep=1pt]{};
    \node at (-2.47,-2.44) [circle,fill,black,inner sep=1pt]{};
    \node at (-2.29,-2.478)[circle,fill,black,inner sep=1pt]{};
    \node at (-1.26,-2.36) [circle,fill,black,inner sep=1pt]{};
    \node at (-1.558,-2.45) [circle,fill,black,inner sep=1pt]{};
    \node at (-.809,-2.106) [circle,fill,black,inner sep=1pt]{};
    \node at (-.69,-2.017) [circle,fill,black,inner sep=1pt]{};
    \node at (-.51,-1.838) [circle,fill,black,inner sep=1pt]{}; 
\end{tikzpicture}    .
\end{equation}
In the illustration above, the blue chords represent matter chords, whereas the black chords correspond to Hamiltonian chords.\\
Now we introduce the left and right ladder operators corresponding to Hamiltonian chords as:
\begin{align} \label{eq:ladder-1}
a_L^{\dagger}\left|n_0, \cdots, n_k\right\rangle&=\left|n_0+1, \cdots, n_k\right\rangle, \quad a_R^{\dagger}\left|n_0, \cdots, n_k\right\rangle=\left|n_0, \cdots, n_k+1\right\rangle, \\
a_L\left|n_0, \cdots, n_k\right\rangle &=\sum_{j=0}^k\left[n_j\right] r_V^j q^{\sum_{l<j} n_l}\left|n_0, \cdots, n_j-1, \cdots, n_k\right\rangle, \\
a_R\left|n_0, \cdots, n_k\right\rangle &=\sum_{j=0}^k\left[n_{k-j}\right] r_V^j q^{\sum_{l>k-j} n_l}\left|n_0, \cdots, n_{k-j}-1, \cdots, n_k\right\rangle.
\end{align}
it is straightforward to show that they satisfy the following commutation relations:
\begin{align}
    [a_{L},a_{L}^{\da}]_{q}	&=[a_{R},a_{R}^{\da}]_{q}=1,\\
[a_{L},a_{R}]	&=[a_{L}^{\da},a_{R}^{\da}]=0, \\
[a_{L},a_{R}^{\da}]	&=[a_{R},a_{L}^{\da}]=r_{V}^{\hat{n}_{M}}q^{\hat{n}_{H}}. \label{eq:def-T1}
\end{align}
where the number operator $\hat{n}_H$ and $\hat{n}_M$ counts the number of Hamiltonian chords and matter chords correspondingly:
\be
\hat{n}_H |n_0,\cdots,n_k\ra =\sum_{i=0}^{k} n_i,\quad \hat{n}_M|n_0,\cdots,n_k\ra =k.
\ee
Similar to \eqref{eq:H0-def}, we introduce the left and right Hamiltonian chord operator as
\be\label{eq:HLR}
H_{L/R}=a_{L/R}+a_{L/R}^{\da}.
\ee
They emerges as the double-scaled limit of Hamiltonian operator in SYK model. For a detailed explanation on this point, we refer the readers to \cite{Lin:2022rbf}. In the current context, $H_L$ and $H_R$ are Hermitian operators that act on the chord Hilbert space $\mathcal{H}$. it is straightforward to examine that the left and right Hamiltonian chord operators commute:
\be
[H_L, H_R] =0.
\ee
We can similarly introduce ladder operators for matter chords as:
\begin{align}\label{eq:ladder-2}
b_{L}^{\da}|n_{0},\cdots,n_{k}\ra&=|0,n_{0},\cdots,n_{k}\ra,\qq b_{R}^{\da}|n_{0},\cdots,n_{k}\ra=|n_{0},\cdots,n_{k},0\ra,\\
    b_{L}|n_{0},\cdots,n_{k}\ra	&=\sum_{j=1}^{k}r^{j-1}r_{V}^{\sum_{l<j}n_{l}}|n_{0},\cdots,n_{j-2},n_{j-1}+n_{j},n_{j+1}\cdots,n_{k}\ra , \\
b_{R}|n_{0},\cdots,n_{k}\ra	&=\sum_{j=1}^{k}r^{j-1}r_{V}^{\sum_{l>k-j}n_{l}}|n_{0},\cdots,n_{k-j-1},n_{k-j}+n_{k-j+1},\cdots,n_{k}\ra,
\end{align}
and they satisfy the following commutation relations:
\begin{align} 
[b_{L},b_{L}^{\da}]_{r}	&=[b_{R},b_{R}^{\da}]_{r}=1, \\
[b_{L},b_{R}]	&=[b_{L}^{\da},b_{R}^{\da}]=0,\\
[b_{L},b_{R}^{\da}]	&=[b_{R},b_{L}^{\da}]=r^{\hat{n}_{M}}r_{V}^{\hat{n}_{H}}. \label{eq:def-T2}
\end{align}
The matter chord operator is defined as:
\be
M_{L/R}=b_{L/R}+b_{L/R}^{\da}.
\ee
and the left and right matter chord operator commutes with each other:
\be \label{eq:MLR}
[M_L, M_R]=0.
\ee
Furthermore, one can show by applying the definition of ladder operators that the left and right generators commute with each other:
\be \label{eq:commuting-LR}
[H_L,H_R]=[H_L, M_R]= [M_L,H_R]=[M_L,M_R]=0.
\ee
We define the left and right chord observable algebras to be the von~Neumann algebras
\be \label{eq:observable-algebras}
\ma_{L/R}:=\operatorname{vN}\left(H_{L/R},M_{L/R}\right).
\ee
More precisely, the unital $*$-algebras of finite polynomials in the generators are weakly dense in $\ma_L$ and $\ma_R$, and their bicommutants give the von~Neumann closures.  This is the double-scaled algebra considered in \cite{Lin:2022rbf}.  This definition includes only the two chord fields $H$ and $M$; in particular, chord-number operators are not adjoined as extra generators.  A typical Wick monomial in the algebraic core of $\ma_L$ has the form
\be \label{eq:basis-1}
H_{L}^{n_{0}}M_L H_{L}^{n_{1}}\cdots M_L H_{L}^{n_{k}}.
\ee
Finite linear combinations of operators of the form \eqref{eq:basis-1} form a weakly dense algebraic core of $\ma_L$, and similarly for $\ma_R$. However, their action on $\Omega$ is complicated, as they generate superposition of states with different amount of chords. To illustrate, let us consider the action of $H_{L}^k$, which yields
\be
H_{L}^k |\Omega\ra = |k\ra + \text{States with amount of chords less than }k.
\ee
it is more convenient in many situations to work with the normal ordered operator basis. Let us consider the situation without matter chords at first. The normal ordered operator basis can be defined recursively as:
\be \label{eq:recursion-Hk}
\normord{H_{L}^{k+1}} = H_L \normord{H_{L}^k}-\wick{\c H_L  
  \normord{\c H_{L}^k}} \equiv H_{L}\normord{H_{L}^{k}}-[k]_{q} \normord{H_{L}^{k-1}},\quad \normord{H_{L}}\equiv H_{L}.
\ee
They generate state with definite amount of Hamiltonian chords:
\be \label{eq:normal-ordered-1}
:H^{k}_L:|\Omega\ra = |k\ra, \quad \forall k \in \mathbb{N}.
\ee
This can be shown by induction. One can assume \eqref{eq:normal-ordered-1} holds for $n\leq k$, and then show it holds for $k+1$ by acting the two sides of \eqref{eq:recursion-Hk} on $\Omega$. Note that the normal-ordering here is different from the conventional one defined by moving all creation operators to the left of all annihilation operators. Here it is defined with respect to the contraction rule of $H_L$ in \eqref{eq:recursion-Hk}. As for matter chord operator $M_L$, we can define the normal ordering in a parallel manner:
\be
:M_{L}^{k+1}:=M_{L}:M_{L}^{k}:-[k]_{r} :M_{L}^{k-1}:,\quad:M_{L}:=M_{L}
\ee
it is straightforward to show 
\be
:M_{L}^{k}:|\Omega\ra=|0,0,\cdots,0\ra, \quad \forall k \in \mathbb{N}.
\ee
We now define the normal ordering for a general operator basis in \eqref{eq:basis-1} as:
\be \label{eq:normal-ordered-def1}
\begin{split}
& \normord{H_L^{n_0+1} M_L H_L^{n_1} \cdots M_L H_L^{n_k}} \equiv H_L \normord{ H_L^{n_0} M_L H_L^{n_1} \cdots M_L H_L^{n_k}} \\
& -\sum_{j=0}^k\left[n_j\right]_q r_V^j q^{\sum_{l<j} n_l} \normord{H_L^{n_0} M_L \cdots M_L H_L^{n_j-1} M_L \cdots M_L H_L^{n_k}},
\end{split}
\ee
where the terms subtracted are all possible contractions between $H_L$ and $H^{n_0}_L M_L\cdots M_L H^{n_k}_L$. For example, we have:
\be
\wick{\c1 H_L\,\normord{H_L^{n_0}M_L\cdots M_L \c1 H_L^{n_j}M_L\cdots } }
=\left[n_j\right]_q r_V^j q^{\sum_{l<j}n_l}
\normord{H_L^{n_0}M_L\cdots M_LH_L^{n_j-1}M_L\cdots}.
\ee
We can apply similar rules to matter chord operators $M_L$, which gives rise to:
\be\label{eq:normal-ordered-def2}
\begin{split}
& \normord{M_L H_L^{n_0} M_L \cdots M_L H_L^{n_k}}\equiv M_L \normord{ H_L^{n_0} M_L \cdots M_L H_L^{n_k}} \\
& -\sum_{j=1}^k r^{j-1} r_V^{\sum_{l<j} n_l} \normord{ H_L^{n_0} \cdots M_L H_L^{n_{j-1}+n_j} M_L \cdots M_L H_L^{n_k}}.
\end{split}
\ee
Equation \eqref{eq:normal-ordered-def1} and \eqref{eq:normal-ordered-def2} completely defines the normal ordering for any strings of operators $M_L$ and $H^{k}_L,\forall k\in \mathbb{N}$. 

Now, let us streamline the notation by introducing the chord field operator $\Phi_L$ as
\be\label{eq:chord-field-L}
\begin{split}
& \Phi_{L}\left(n_{0},\cdots,n_{k}\right)\equiv \normord{H_{L}^{n_{0}}M_{L}H_{L}^{n_{1}}\cdots H_{L}^{n_{k-1}}M_{L}H_{L}^{n_{k}}},\\
&\Phi_{L}\left(\Omega\right)\equiv\mathbf{1},\qq\Phi_{L}\left(0,0,\cdots,0\right)=\normord{M^{k}},\qq\Phi\left(k\right)=\normord{H^k}.
\end{split}
\ee
One can show following the same strategy as above that a general state $|n_0,\cdots, n_k\ra $ can be generated by acting $\Phi_L (n_0,\cdots,n_k)$ on the empty state $\Omega$:
\be \label{eq:phi-L}
\Phi_{L}\left(n_{0},\cdots,n_{k}\right)|\Omega\ra=|n_{0},\cdots,n_{k}\ra.
\ee
We can draw parallels with conventional quantum field theory by incorporating a classical configuration space of chords into the framework:
\be
\mathcal{C}:= \cup_{k=0}^{\infty}\left\{\left(n_0, \cdots, n_k\right) \in \mathbb{N}^{k+1}\right\},\quad \Omega \equiv (0).
\ee
Then $\Phi_L: \mathcal{C}\to \ma_L$ is an operator-valued distribution on $\mathcal{C}$. For a given field configuration $x=(n_0,\cdots,n_k)\in\mathcal{C}$ , $\Phi_L (x)$ is a field operator that generates a state $|x\ra=|n_0,\cdots,n_k\ra\in\mathcal{H}$ from the empty state $\Omega$. By construction, the finite linear span of $\Phi_L(x)$ is an algebraic Wick core for $\ma_L$, while the vectors $\Phi_L(x)|\Omega\ra$ span a dense subspace of $\mh$. Its equivalence to the original basis in \eqref{eq:basis-1} can be verified by the following observation:

We introduce an ordering on the set of operator basis elements in  
\eqref{eq:basis-1} by comparing tuples of the form \((n_0, n_1, \dots, n_k)\), where \(k\) denotes the number of matter 
insertions \(M\) in the operator monomial, and \((n_0, \dots, n_k)\) counts the number of Hamiltonians \(H\) appearing between the matter chords.  We say $\Phi(n_0,n_1,\dots, n_k)$ is of lower order to $\Phi(m_0,m_1,\dots,m_{k^\pp})$ if either \(k < k^\pp\), or \(k = k^\pp \) and the sequence \((m_0, \dots, m_{k})\) is lexicographically greater than \((n_0, \dots, n_k)\).
With this definition, the left-hand sides of~\eqref{eq:normal-ordered-def1} and~\eqref{eq:normal-ordered-def2} are equal to the respective leading terms on the right-hand sides, while the terms subtracted off are of lower order. By induction on the operator order one can deduce that the normal ordered basis $\{\Phi_L (x),\forall x\in \mathcal{C}\}$ is equivalent to the original basis \eqref{eq:basis-1}.

One can introduce the normal ordered field operator for the right operators in $\ma_R$. We list corresponding definitions as follows:
\be
\begin{split}
    &\normord{H_{R}^{k+1}}\equiv H_R \normord{H_{R}^k}- [k]_q \normord{H^{k-1}_R}, \quad \normord{H_R} \equiv H_R,  \\
    &\normord{M_{R}^{k+1}}\equiv M_R \normord{M_{R}^k}- [k]_r \normord{M^{k-1}_R}, \quad \normord{M_R} \equiv M_R.
\end{split}
\ee
and:
\be \label{eq:normal-ordered-def3}
\begin{split}
& \normord{H_{R}^{n_0+1} M_R H_R^{n_1} \cdots M_R H_R^{n_k}} \equiv H_R \normord{ H_R^{n_0} M_R H_R^{n_1} \cdots M_R H_R^{n_k}} \\
& -\sum_{j=0}^k\left[n_{k-j}\right]_q r_V^j q^{\sum_{l>k-j} n_l} \normord{H_R^{n_0} M_R \cdots M_R H_R^{n_{k-j}-1} M_R \cdots M_R H_R^{n_k}}.
\end{split}
\ee
\be\label{eq:normal-ordered-def4}
\begin{split}
& \normord{M_R H_R^{n_0} M_R \cdots M_R H_R^{n_k}}\equiv M_R \normord{ H_R^{n_0} M_R \cdots M_R H_R^{n_k}} \\
& -\sum_{j=1}^k r^{j-1} r_V^{\sum_{l>k-j} n_l} \normord{ H_R^{n_0} \cdots M_R H_R^{n_{k-j}+n_{k-j+1}} M_R \cdots M_R H_{R}^{n_k}}.
\end{split}
\ee
The right chord field operator $\Phi_R: \mathcal{C}\to \ma_R$ is then defined as
\be
\Phi_R (n_0,\cdots,n_k) \equiv \normord{H_{R}^{n_0} M_R H_R^{n_1} \cdots M_R H_R^{n_k}}.
\ee
Different from \eqref{eq:phi-L}, it generates a state with reversed ordering from empty state:
\be\label{eq:phi-R}
\Phi_{R}\left(n_{0},\cdots,n_{k}\right)|\Omega\ra=|n_{k},\cdots,n_{0}\ra.
\ee
In conclusion, the algebras $\mathcal A_{L/R}$ are generated by chord fields $\Phi_{L/R}(x)$ associated with configurations $x\in\mathcal C$.  The operator--state correspondence gives
\be \label{eq:op-state-correspond}
|n_0,\cdots,n_k\rangle
=\Phi_L(n_0,\dots,n_k)|\Omega\rangle
=\Phi_R(n_k,\dots,n_0)|\Omega\rangle,
\ee
so $|\Omega\ra$ is cyclic for both algebras.

The same relations give separatingness.  If $A_L\in\ma_L$ and $A_L|\Omega\ra=0$, then for every $B_R\in\ma_R$,
\be
A_LB_R|\Omega\ra=B_RA_L|\Omega\ra=0,
\ee
where we used~\eqref{eq:commuting-LR}.  Since $\ma_R|\Omega\ra$ is dense by the right operator--state correspondence, $A_L=0$.  Interchanging left and right gives the same conclusion for $\ma_R$.  Thus $|\Omega\ra$ is cyclic and separating for both double-scaled algebras.

\subsection{Exploring Modular Structure of the Double-Scaled Algebra} \label{sec:Modular}
We now explore the modular structure and show that the left and right algebras are mutual commutants.  On the dense Wick core $\ma_L^{\rm alg}|\Omega\ra$, define the conjugate-linear Tomita map~\cite{Witten_2018} by
\be
S_0\Psi|\Omega\ra=\Psi^{\da}|\Omega\ra,
\qquad \Psi\in\ma_L^{\rm alg}.
\ee
For a chord word $x=(n_0,\ldots,n_k)$,
\be \label{eq:state-x}
\Phi_L(x)|\Omega\ra=|x\ra=|n_0,\ldots,n_k\ra,
\ee
and the Wick recursion gives
\be
\Phi_L(n_0,\ldots,n_k)^\dagger=\Phi_L(n_k,\ldots,n_0).
\ee
Hence
\be \label{eq:Tomita-action}
S_0|n_0,\ldots,n_k\ra=|n_k,\ldots,n_0\ra.
\ee
For a finite linear combination, the coefficients are complex conjugated.  Reflection of a sewn chord diagram preserves every crossing weight and exchanges bra and ket, so
\be \label{eq:reflection-antiunitary}
\la S_0\xi|S_0\eta\ra=\la\eta|\xi\ra
\qquad (\xi,\eta\in\ma_L^{\rm alg}|\Omega\ra).
\ee
Thus $S_0$ extends to an anti-unitary involution $J$ on $\mh$.  Its closure is the Tomita operator $S_\Omega=J$, and the polar decomposition gives
\be
S_\Omega=J\Delta_\Omega^{1/2},
\qquad
\Delta_\Omega=S_\Omega^\dagger S_\Omega=\mathbf1.
\ee
This reversal agrees with the reflection operator $\mathrm R$ of~\cite{Lin_2023}.

Now let us prove that the left and right algebras are commutants of each other. It is straightforward to show that $\ma_L \subseteq \ma^{\pp}_R$ and $\ma_R \subseteq \ma^{\pp}_{L}$ by the fact that the generators of the two algebras commute with each other, as shown in \eqref{eq:commuting-LR}.  A direct application of Tomita-Takesaki theory shows that $J\ma_L J = \ma_{L}^\pp$. Hence, to prove the equivalence between $\ma_R$ and $\ma^{\prime}_L$,  we only need to show $J \ma_L J \subseteq \ma_R$. This can be verified by examining the action of $J H_L J$ and $J M_L J$ on a generic state $|n_0,\cdots, n_k\ra \in \mh$.
we find
\be
\begin{split}
JH_{L}J|n_{0},\cdots, & n_{k}\ra	=J H_{L}|n_{k},\cdots,n_{0}\ra \\
	&=J\left(|n_{k}+1,\cdots,n_{0}\ra+\sum_{j=0}^{k}[n_{k- j}]r_{V}^{j}q^{\sum_{l>k-j}n_{l}}|n_{k},\cdots,n_{j}-1,\cdots,n_{0}\ra\right) \\
	&=|n_{0},\cdots,n_{k}+1\ra+\sum_{j=0}^{k}[n_{k-j}]r_{V}^{j}q^{\sum_{l>k-j}n_{l}}|n_{0},\cdots,n_{k-j}-1,\cdots,n_{k}\ra \\
	&=H_{R}|n_{0},\cdots,n_{k}\ra,
\end{split}
\ee
which implies that $J H_L J = H_R$. Similarly, one can verify that $J M_L J=M_R$. These relations among generators can then be extended to a dense subset of $\ma_L$ consisting of finite linear span of strings of $H_L$ and $M_L$. By taking closure of the subset, this confirms that $J \ma_L J \subseteq \ma_R$. We conclude that $\ma_L$ and $\ma_R$ are indeed commutants of each other, with the following relations satisfied: 
\be
\ma^{\pp}_L = \ma_R,\quad \ma^{\pp}_R = \ma_L. 
\ee

\subsection{Tracial Property of \texorpdfstring{$\Omega$}{Omega} and the Type of the Double-Scaled Algebra}
\label{sec:proof}

We now determine the trace and the von~Neumann type of the double-scaled algebra.  In this
subsection we write $\ma\equiv\ma_L$ and use the result of the previous subsection,
\be
\ma^{\pp}=\ma_R.
\ee
We work in the regime
\be \label{eq:qstar}
\rho:=\max\{|q|,|r|,|r_V|\}<1.
\ee
In this regime the chord creation and annihilation operators are bounded, as we show below using a $Q$-twisted Fock Hilbert space.  We first show that the empty chord state
defines a normalized trace on $\ma$. We then revisit the finite-support argument for the center used in the original version of this paper. The argument is complete for vectors with finite chord-number support. The only additional ingredient needed for a general central element $Z\in\ma$ is an estimate showing that the possible infinite chord-number tail of $Z|\Omega\rangle$ cannot evade the finite-support obstruction. Together these results imply that $\ma$ is a Type~II$_1$ factor.

\paragraph{Traciality of the empty chord state.}
In sections~\ref{sec:construction} and~\ref{sec:Modular} we showed that $|\Omega\ra$ is
cyclic and separating for $\ma$, and that its Tomita operator is the anti-linear
chord-reversal operator,
\be
S_\Omega|n_0,\ldots,n_k\ra=|n_k,\ldots,n_0\ra.
\ee
Since chord reversal preserves the inner product, $S_\Omega$ is anti-unitary.  Its
polar decomposition therefore gives
\be
\Delta_\Omega=\mathbf 1.
\ee
The modular automorphism group is trivial, and the empty-state expectation value
\be \label{eq:vacuum-trace}
\tau(a):=\la\Omega|a|\Omega\ra,
\qquad \forall a \in\ma,
\ee
is therefore tracial:
\be
\tau(AB)=\tau(BA),
\qquad A,B\in\ma.
\ee

The remaining properties follow directly in the standard representation on $\mh$.
The functional $\tau$ is normal because it is the restriction of a vector functional
on $\mathcal B(\mh)$.  More explicitly, if $0\leq A_\alpha\uparrow A$ in $\ma$, then
$A_\alpha\to A$ strongly and
\be
\tau(A)=\lim_\alpha\tau(A_\alpha)=\sup_\alpha\tau(A_\alpha).
\ee
It is faithful because $|\Omega\ra$ is separating: if $\tau(A^{\da}A)=0$, then
$A|\Omega\ra=0$, and hence $A=0$.  Finally,
\be
\tau(\mathbf 1)=1.
\ee
Thus $\tau$ is a faithful normal finite tracial state.  In particular, no separate
energy-window argument is needed to establish semifiniteness.

\paragraph{$Q$-twisted Fock Hilbert space realization.}
For the factoriality argument, it is useful to identify the chord Hilbert space with a
$Q$-twisted Fock Hilbert space~\cite{Boejko_1997}.  Besides making properties of the chord inner
product more transparent, this realization gives a uniform operator-norm bound for the
creation and annihilation operators.

Introduce the species-dependent crossing matrix
\be \label{eq:mixed-Q-matrix}
Q_{HH}=q,
\qquad Q_{MM}=r,
\qquad Q_{HM}=Q_{MH}=r_V.
\ee
We write the one-sided chord fields as
\be
H=a_H+a_H^{\da},
\qquad
M=a_M+a_M^{\da},
\ee
where $a_H$ and $a_M$ replace the earlier notation $a$ and $b$.  The one-sided ladder
operators satisfy the twisted commutation relations
\be \label{eq:commutator-one-side}
a_Ia_J^{\da}-Q_{IJ}a_J^{\da}a_I=\delta_{IJ},
\qquad I,J\in\{H,M\},
\ee
with no summation over $I,J$. Let
\be
\mathfrak h=\mathbb C e_H\oplus\mathbb C e_M,
\qquad
\la e_I|e_J\ra=\delta_{IJ},
\ee
be the two-dimensional Hilbert space of chord species, and define the twist operator
\be \label{eq:mixed-Q-YB-operator}
T(e_I\otimes e_J)=Q_{IJ}\,e_J\otimes e_I,
\qquad I,J\in\{H,M\}.
\ee
Since $Q$ is real and symmetric, $T$ is self-adjoint, and
\be
\|T\|=\rho<1.
\ee
Now consider the space $\mathfrak h^{\otimes N}$ of $N$-amount of chords, let $T_i$ act as $T$ on the $i$th and $(i+1)$st
factors and as the identity on the remaining factors.  These operators obey
\be
T_iT_{i+1}T_i=T_{i+1}T_iT_{i+1},
\qquad
T_iT_j=T_jT_i\quad (|i-j|\geq2).
\ee
Thus, if $\sigma=s_{i_1}\cdots s_{i_\ell}\in S_N$ is a reduced decomposition into
adjacent transpositions, the operator
\be
\varphi_N(\sigma):=T_{i_1}\cdots T_{i_\ell}
\ee
is independent of the chosen reduced decomposition.  The corresponding
$Q$-symmetrizer is
\be \label{eq:mixed-Q-symmetrizer}
P_Q^{(N)}:=\sum_{\sigma\in S_N}\varphi_N(\sigma),
\qquad
P_Q^{(0)}:=1.
\ee
For example,
\be
\begin{split}
P_Q^{(3)}(e_H\otimes e_M\otimes e_H)
={}&(1+r_V^2q)e_H\otimes e_M\otimes e_H \\
&+r_V(1+q)e_M\otimes e_H\otimes e_H
+r_V(1+q)e_H\otimes e_H\otimes e_M.
\end{split}
\ee
Start from the standard Fock Hilbert space space
\be
\mathcal F (\mathfrak h)
=\mathbb C\Omega_Q\oplus\bigoplus_{N\geq1}^{\rm alg}\mathfrak h^{\otimes N}
\ee
and equip its $N$-particle sector with the $Q$-deformed inner product
\be \label{eq:mixed-Q-inner-product}
\la\xi|\eta\ra_Q
:=\la\xi|P_Q^{(N)}\eta\ra_0,
\qquad
\xi,\eta\in\mathfrak h^{\otimes N},
\ee
where $\la\cdot|\cdot\ra_0$ is the standard tensor-product inner product and
$\la\Omega_Q|\Omega_Q\ra_Q=1$.  For words $v=v_1\cdots v_N$ and
$w=w_1\cdots w_N$ in the alphabet $\{H,M\}$, this inner product is
\be \label{eq:mixed-Q-inner-product-words}
\left\langle e_{v_1}\otimes\cdots\otimes e_{v_N},
 e_{w_1}\otimes\cdots\otimes e_{w_N}\right\rangle_Q  \, =
\sum_{\sigma\in S_N}
\left(\prod_{a=1}^{N}\delta_{v_a,w_{\sigma^{-1}(a)}}\right)
\left(\prod_{\substack{1\leq a<b\leq N\\ \sigma(a)>\sigma(b)}}
Q_{w_aw_b}\right).
\ee
A permutation contributes only when it matches equal chord species, and each inversion
contributes the crossing weight of the two species that cross.

For $\rho<1$, the operator $P_Q^{(N)}$ is positive and has trivial kernel for every
$N$~\cite{jorgensen2000}.  Hence \eqref{eq:mixed-Q-inner-product} is a positive,
non-degenerate inner product~\cite{Boejko_1997,skalski2017}.  We denote the Hilbert
space completion by $\mathcal F_Q(\mathfrak h)$ and refer to it as the $Q$-twisted
Fock Hilbert space.

We next compare this construction with the chord representation.  For a word
$w=w_1\cdots w_N$, with $w_j\in\{H,M\}$, define
\be
\Phi_L(w):=\normord{w_{1,L}w_{2,L}\cdots w_{N,L}},
\qquad
|w\ra:=\Phi_L(w)|\Omega\ra.
\ee
The trace defines the $L^2$ inner product
\be
\la\Phi|\Psi\ra_{2,\tau}:=\tau(\Phi^{\da}\Psi).
\ee
Sewing the lower endpoints of $\Phi_L(v)^{\da}$ to the upper endpoints of
$\Phi_L(w)$ produces a permutation of the open chords.  Species must agree along each
sewn chord, while an intersection between species $I$ and $J$ contributes $Q_{IJ}$.
Therefore
\be \label{eq:chord-Q-isometry}
\la\Phi_L(v)|\Phi_L(w)\ra_{2,\tau}
=\la v|w\ra
=\left\langle e_{v_1}\otimes\cdots\otimes e_{v_N},
 e_{w_1}\otimes\cdots\otimes e_{w_N}\right\rangle_Q,
\ee
and both sides vanish when the two words have different lengths.

Let $\Phi_L$ be the linear span of the finite normal-ordered words.  The map
\be \label{eq:mixed-Q-identification}
\mathsf U_0:\quad
\Phi_L(w)\longmapsto e_{w_1}\otimes\cdots\otimes e_{w_N},
\qquad
\mathbf 1\longmapsto\Omega_Q,
\ee
is therefore an isometry from $\mathcal P_L$, equipped with the $L^2$ norm, onto the  Fock space.  Since $\mathcal P_L$ is dense in $L^2(\ma_L,\tau)$ and the
elementary tensors are dense in $\mathcal F_Q(\mathfrak h)$, it extends uniquely to a
unitary
\be \label{eq:mixed-Q-unitary}
\mathsf U:L^2(\ma_L,\tau)\longrightarrow\mathcal F_Q(\mathfrak h).
\ee
After the operator-state correspondence $A\mapsto A|\Omega\ra$, this is precisely a
unitary identification of the chord Hilbert space $\mh$ with
$\mathcal F_Q(\mathfrak h)$.  In particular,due to triviality of the kernel of the $Q$-symmetrizer, the unitary equivalence shows that the fixed-length normalized chord operators have a
non-degenerate Gram matrix and form a basis of each chord-number sector.

The unitary also intertwines the chord ladder operators with the $Q$-twisted Fock
operators.  Let left creation be
\be
\ell_I^{\da}\xi=e_I\otimes\xi.
\ee
Its adjoint with respect to \eqref{eq:mixed-Q-inner-product} is the annihilation operator
\be \label{eq:mixed-Q-annihilation}
\ell_I(e_{w_1}\otimes\cdots\otimes e_{w_N})
=
\sum_{a=1}^{N}\delta_{I,w_a}
\left(\prod_{b<a}Q_{Iw_b}\right)
 e_{w_1}\otimes\cdots\otimes\widehat{e_{w_a}}\otimes\cdots\otimes e_{w_N}.
\ee
This is exactly the contraction rule appearing in the recursive definition of the
normal-ordered chord operators.  Thus, for $I=H,M$,
\be \label{eq:mixed-Q-intertwining}
\mathsf U a_I\mathsf U^{-1}=\ell_I,
\qquad
\mathsf U a_I^{\da}\mathsf U^{-1}=\ell_I^{\da}.
\ee
This realization also gives the required operator-norm bounds.  For
$\xi\in\mathfrak h^{\otimes N}$,
\be
\|\ell_I^{\da}\xi\|_Q^2
=
\la e_I\otimes\xi|P_Q^{(N+1)}(e_I\otimes\xi)\ra_0.
\ee
In appendix~\ref{app:Bound-PQ} we prove the operator inequality
\be
P_Q^{(N+1)}
\leq
\frac{1}{1-\rho}\bigl(\mathbf 1\otimes P_Q^{(N)}\bigr).
\ee
It follows that
\be
\|\ell_I^{\da}\xi\|_Q^2 \leq
\frac{1}{1-\rho}
\la e_I\otimes\xi|
(\mathbf 1\otimes P_Q^{(N)})(e_I\otimes\xi)\ra_0 =
\frac{1}{1-\rho}\|\xi\|_Q^2.
\ee
The bound is uniform in $N$, and the chord-number sectors are mutually orthogonal.
Therefore $\ell_I^{\da}$ extends to a bounded operator on the full Fock Hilbert space,
and
\be \label{eq:bound-ell}
\|\ell_I^{\da}\|=\|\ell_I\|
\leq\frac{1}{\sqrt{1-\rho}}.
\ee
By the unitary equivalence \eqref{eq:mixed-Q-intertwining}, the chord ladder operators
obey the same bound:
\be \label{eq:bound-a}
\|a_H\|,\ \|a_H^{\da}\|,\ \|a_M\|,\ \|a_M^{\da}\|
\leq\frac{1}{\sqrt{1-\rho}}.
\ee
These uniform bounds will be used below to control the infinite chord-number tail.

\paragraph{Factoriality on the finite chord-number subspace.}
We first isolate the simple finite support argument behind factoriality.  For
$x=(n_0,\ldots,n_k)$, its total chord length is defined by
\be
|x|:=n_0+\cdots+n_k+k.
\ee
Let $\Pi_N$ be the projection onto the finite-dimensional subspace $\mh^{(N)}$ spanned
by chord words of length $N$.  By the $Q$-twisted Fock realization, these words form a
basis of $\mh^{(N)}$.  We denote a word in the two letters $H$ and $M$ by $w$; the
symbols $Hw,Mw,wH,wM$ denote left and right concatenation.

Let $Z\in\mathcal Z(\ma)$ and set
\be
|\xi\ra:=\bigl(Z-\tau(Z)\mathbf 1\bigr)|\Omega\ra.
\ee
For $X\in\{H,M\}$, centrality of $Z$, commutativity of the left and right algebras,
and $X_L|\Omega\ra=X_R|\Omega\ra$ imply
\be \label{eq:central-vector-equations}
(X_L-X_R)|\xi\ra=0.
\ee
Suppose first that $|\xi\ra$ has finite chord-length support.  Since
$\la\Omega|\xi\ra=0$, a nonzero $|\xi\ra$ has a largest occupied length $N\geq1$.
Write $|\xi_N\ra:=\Pi_N|\xi\ra$.  The length-$(N+1)$ part of
\eqref{eq:central-vector-equations} receives contributions only from the creation
operators, and hence
\be
D_H^{(N)}|\xi_N\ra=D_M^{(N)}|\xi_N\ra=0,
\ee
where
\be \label{eq:leading-maps}
D_H^{(N)}|w\ra:=|Hw\ra-|wH\ra,
\qquad
D_M^{(N)}|w\ra:=|Mw\ra-|wM\ra.
\ee
The two maps have no common kernel for $N\geq1$.  To see why this is the case, let
 $|\eta\ra=\sum_{|w|=N}c_w|w\ra$.  If $w$ ends in $H$, the coefficient of $|Mw\ra$ in
$D_M^{(N)}|\eta\ra$ is exactly $c_w$, since every word of the form $w'M$ ends in $M$.
If $w$ ends in $M$, the coefficient of $|Hw\ra$ in $D_H^{(N)}|\eta\ra$ is exactly
$c_w$.  Thus
$D_H^{(N)}|\eta\ra=D_M^{(N)}|\eta\ra=0$ forces all $c_w$ to vanish, in contradiction
with $|\xi_N\ra\neq0$.

It follows that a central $Z$ is a constant operator whenever $Z|\Omega\ra$ has finite chord-length
support.  This is essentially the finite length support  proof in previous version of the paper.  Its
limitation is now explicit: for a general bounded operator $Z\in\ma$, the vector
$Z|\Omega\ra$ may have an infinite chord-number tail and therefore need not possess a
largest occupied length.  We now show that this tail cannot evade the finite-length
obstruction.

\paragraph{Controlling the infinite chord-number tail.}
The remainder of the proof follows the strategy of~\cite{skalski2017}.  Define the
single-species subspaces
\be
\mh_H:=\overline{\operatorname{span}\{|n\ra:n\geq0\}},
\qquad
\mh_M:=\overline{\operatorname{span}\{|M^k\ra:k\geq0\}},
\ee
where $|M^k\ra$ denotes the state with $k$ matter chords and no Hamiltonian chords.
These are the pure Hamiltonian- and pure matter-chord subspaces.  We also define
\be
\mb_H:=\operatorname{vN}(H_L),
\qquad
\mb_M:=\operatorname{vN}(M_L).
\ee
The vacuum spectral distributions of $H_L$ and $M_L$ are the $q$- and
$r$-semicircular distributions, respectively~\cite{Boejko_1997}.  For
$|q|,|r|<1$, both have continuous densities and therefore no delta-function weight at
any individual eigenvalue.  More precisely, if $E_H(B)$ is the spectral projection of
$H_L$ associated with a Borel set $B\subset\mathbb R$, then
\be
\mu_H(B):=\tau(E_H(B)),
\qquad
\mu_H(\{\lambda\})=0
\quad\text{for all }\lambda\in\mathbb R,
\ee
and similarly for $M_L$.  Thus the abelian algebras $\mb_H$ and $\mb_M$ contain no
nonzero minimal projections; equivalently, they are diffuse.

The operator--state correspondence restricted to these algebras gives
\be
\overline{\mb_H|\Omega\ra}=\mh_H,
\qquad
\overline{\mb_M|\Omega\ra}=\mh_M.
\ee
Diffuseness allows us to choose, for each $I=H,M$, self-adjoint unitaries
$u_\alpha^{(I)}\in\mb_I$ such that
\be \label{eq:weak-unitaries}
\bigl(u_\alpha^{(I)}\bigr)^2=\mathbf 1,
\qquad
|\zeta_\alpha^{(I)}\ra:=u_\alpha^{(I)}|\Omega\ra
\longrightarrow0
\quad\text{weakly in }\mh_I.
\ee
For concreteness, this sequence may be constructed explicitly.  A diffuse abelian
finite von~Neumann algebra with trace $\tau$ can be represented as
\be
(A,\tau)\simeq
\left(L^\infty(X,\gamma),\ \int_X\!\cdot\,\dd\gamma\right),
\ee
where the probability measure has no delta-function component.  Choose a nested dyadic
partition
\be
X=\bigsqcup_{\varepsilon\in\{0,1\}^n}E_\varepsilon,
\qquad
\gamma(E_\varepsilon)=2^{-n},
\ee
and define the Rademacher functions
\be
u_n:=\sum_{\varepsilon\in\{0,1\}^n}
(-1)^{\varepsilon_n}\mathbf 1_{E_\varepsilon}.
\ee
Here $\varepsilon_n$ is the last binary digit of $\varepsilon$.  Each $u_n$ is a
self-adjoint unitary, and the nested partitions at different $n$ imply 
\be
\tau(u_m u_n)=\delta_{mn}.
\ee
In the GNS representation the vacuum corresponds to the constant function $1$, so
$\{u_n|\Omega\ra\}$ is an orthonormal sequence. For
every $|\chi\ra\in L^2(X,\gamma)$,  Bessel's inequality shows, 
\be
\sum_{n=1}^{\infty}|\la\chi|u_n\Omega\ra|^2\leq\|\chi\|^2,
\ee
and hence $u_n|\Omega\ra$ converges to zero weakly.  This proves the existence of weakly-vanishing sequence of unitaries
\eqref{eq:weak-unitaries} within diffuse abelian finite von Neumann algebras.

\paragraph{An overlap estimate for finite ladder-operator strings.}

Introduce uniform notation for the left and right annihilation operators,
\be
a_{H,L}=a_L,
\quad a_{M,L}=b_L,
\qquad
a_{H,R}=a_R,
\quad a_{M,R}=b_R,
\ee
with $a_{I,L/R}^{\da}$ denoting the corresponding creation operators.  All
left--right ladder commutators vanish except when an annihilation operator meets a
creation operator of the same species.  On the total chord-number sector $\mh^{(N)}$,
the two-sided commutation relations give
\be \label{eq:commutator-decay}
\left(a_{I,L}a_{J,R}^{\da}-a_{J,R}^{\da}a_{I,L}\right)\Pi_N
=\delta_{IJ}\mathcal T_I^{(N)},
\qquad
\|\mathcal T_I^{(N)}\|\leq\rho^N.
\ee
Here $\mathcal T_H$ and $\mathcal T_M$ are the diagonal chord-number operators appearing
on the right-hand sides of \eqref{eq:def-T1} and \eqref{eq:def-T2}. Only the norm bound
will be needed below.

Let $\mh_I$ be one of the two single-species subspaces.  It decomposes as
\be
\mh_I=\bigoplus_{N\geq0}\mh_{I,N},
\qquad
\mh_{I,N}:=\mh_I\cap\mh^{(N)},
\ee
where each $\mh_{I,N}$ is one-dimensional.  Let
\be
A:=a_s\cdots a_1,
\qquad
B:=b_1\cdots b_t,
\ee
where $a_1,\ldots,a_s$ are fixed annihilation operators at one boundary and
$b_1,\ldots,b_t$ are fixed ladder operators at the opposite boundary.  Assume
\be
A\mh_I=0.
\ee
In the application below this is realized by arranging
$a_j\mh_I\subset\mh_I$ for $j<s$ and choosing $a_s$ to annihilate the opposite chord
species, so that $a_s\mh_I=0$. For example, if $I$ is $H$, then we choose $a_s$ to be the annihilation operator of $M$. 

Suppose that $|\chi_\alpha\ra$ is bounded in $\mh$, while
$|\zeta_\alpha\ra\in\mh_I$ is bounded and converges weakly to zero.  We claim that
\be \label{eq:finite-string-mixing}
\lim_\alpha
\la a_1^{\da}\cdots a_s^{\da}\chi_\alpha
\bigm|b_1\cdots b_t\zeta_\alpha\ra=0.
\ee
Indeed,
\be
\la a_1^{\da}\cdots a_s^{\da}\chi_\alpha|B\zeta_\alpha\ra
=
\la\chi_\alpha|AB\zeta_\alpha\ra.
\ee
We can move all $a_i$s to the right of $B$, with 
\be
A B
=
B A
+
\sum_{i=1}^s a_s\cdots a_{i+1}[a_i,B]a_{i-1}\cdots a_1,
\ee
note that 
\be
[a_i,B]
=
\sum_{j=1}^t
b_1\cdots b_{j-1}[a_i,b_j]b_{j+1}\cdots b_t.
\ee
Therefore
\be
A B
={}B A  + \sum_{i=1}^s\sum_{j=1}^t
 a_s\cdots a_{i+1}
 b_1\cdots b_{j-1}[a_i,b_j]
 b_{j+1}\cdots b_t
 a_{i-1}\cdots a_1.
\ee
The first term vanishes on $\mh_I$ because $A\mh_I=0$.  Denote the sum of the
remaining terms by $R$.  We now show that, for any fixed $\kappa$ satisfying
\be
\rho<\kappa<1,
\ee
there is a constant $C$ independent of $N$ such that
\be \label{eq:C-bound}
\|R\Pi_N\|\leq C\kappa^N.
\ee

Consider one term $S_1[a_i,b_j]S_2$, where
\be
S_1:=a_s\cdots a_{i+1}b_1\cdots b_{j-1},
\qquad
S_2:=b_{j+1}\cdots b_ta_{i-1}\cdots a_1.
\ee
If $D:=|S_2|=t-j+i-1$, then the fixed string $S_2$ changes total chord number by at
most $D$.  Writing
\be
\Pi_{[m_-,m_+]}:=\sum_{m=m_-}^{m_+}\Pi_m,
\ee
we have
\be
S_2\Pi_N
=
\Pi_{[\max\{N-D,0\},N+D]}S_2\Pi_N.
\ee
The commutator $[a_i,b_j]$ is either zero or one of the diagonal operators in
\eqref{eq:commutator-decay}.  Hence, for $N\geq D$,
\be
\begin{split}
\|S_1[a_i,b_j]S_2\Pi_N\|
&\leq
\|S_1\|
\max_{|m-N|\leq D}\rho^m\,
\|S_2\Pi_N\| \\
&\leq
\kappa^{-D}\|S_1\|\,\|S_2\|\,\kappa^N.
\end{split}
\ee
The finitely many sectors $N<D$ can be absorbed into the constant.  Moreover, every
factor in $S_1$ and $S_2$ is bounded by \eqref{eq:bound-a}, and therefore
\be \label{eq:bound-S}
\|S_2\Pi_N\|\leq\|S_2\|,
\qquad
\|S_\nu\|\leq(1-\rho)^{-|S_\nu|/2},
\qquad \nu=1,2.
\ee
Since the double sum contains only finitely many terms, this proves
\eqref{eq:C-bound}.

Using $A\zeta_\alpha=0$ and decomposing $\zeta_\alpha$ into chord-number sectors, we
obtain
\be \label{eq:tail-sum}
\left|
\la a_1^{\da}\cdots a_s^{\da}\chi_\alpha
\bigm|b_1\cdots b_t\zeta_\alpha\ra
\right|
\leq
C\sup_\alpha\|\chi_\alpha\|
\sum_{N\geq0}\kappa^N\|\Pi_N\zeta_\alpha\|.
\ee
For fixed $K$, the part with $N\leq K$ tends to zero: weak convergence implies norm
convergence after projection onto the finite-dimensional space
$\bigoplus_{N=0}^K\mh_{I,N}$.  The infinite length tail is uniformly bounded by
\be
\sum_{N>K}\kappa^N\|\Pi_N\zeta_\alpha\|
\leq
\left(\sum_{N>K}\kappa^{2N}\right)^{1/2}
\|\zeta_\alpha\|,
\ee
which tends to zero uniformly in $\alpha$ as $K\to\infty$.  This proves
\eqref{eq:finite-string-mixing}.  The  factor $\kappa^N$ is precisely what
prevents an uncontrolled contribution from the infinite chord-number tail.

\paragraph{Finite chord operator outside a pure-chord sector.}
We now apply \eqref{eq:finite-string-mixing} to the right operator associated with a
finite chord vector.  Here ``finite'' means finite chord-number support:
\be
|\eta\ra\in\bigoplus_{N=0}^{N_{\rm max}}\mh^{(N)}
\ee
for some $N_{\rm max}<\infty$.  Let $|\eta\ra\perp\mh_I$ and let
$\Phi_R(\eta)\in\ma_R$ be the finite right normal-ordered operator satisfying
\be
\Phi_R(\eta)|\Omega\ra=|\eta\ra.
\ee
We claim that, for every $Y\in\ma$, and a weakly vanishing sequence of unitaries $u^{(I)}_\alpha$ defined in \eqref{eq:weak-unitaries}, 
\be \label{eq:mixing-estimate}
\lim_\alpha
\la\Phi_R(\eta)u_\alpha^{(I)}\Omega
\bigm|
Y u_\alpha^{(I)}\Omega\ra=0.
\ee

First take $Y$ to be a finite Wick polynomial.  By linearity, it is enough to take
$|\eta\ra$ to be a single chord word containing at least one letter different from
$I$.  The Wick expansion of $\Phi_R(\eta)$ is a finite sum of normal-ordered ladder
monomials, and every letter of the original chord word appears in each monomial as
either a creation or an annihilation operator.  Hence each monomial contains at least
one ladder operator whose species differs from $I$.  A typical term has the form
\be
R_0
=N_{R_0}
 a_{i_1,R}^{\da}\cdots a_{i_s,R}^{\da}
 a_{i_{s+1},R}\cdots a_{i_p,R},
\ee
where at least one $i_k$ differs from $I$.  Let
\be
s^{\pp}:=\min\{k:i_k\neq I\}.
\ee
If $s^{\pp}>s$, then the annihilation string contains an operator of the wrong species
and therefore annihilates every vector in $\mh_I$; hence
$R_0|\zeta_\alpha^{(I)}\ra=0$.  If $s^{\pp}\leq s$, write
\be
R_0|\zeta_\alpha^{(I)}\ra
=
N_{R_0}
 a_{i_1,R}^{\da}\cdots a_{i_{s^{\pp}},R}^{\da}
|\chi_\alpha\ra,
\ee
where $|\chi_\alpha\ra$ is obtained by applying the remaining fixed ladder operators
to $|\zeta_\alpha^{(I)}\ra$.  The sequence $|\chi_\alpha\ra$ is uniformly bounded by
\eqref{eq:bound-a}.  By the minimality of $s^{\pp}$,
\be
a_{i_j,R}\mh_I\subset\mh_I
\quad (j<s^{\pp}),
\qquad
a_{i_{s^{\pp}},R}\mh_I=0.
\ee
Thus \eqref{eq:finite-string-mixing}, with the opposite-boundary string chosen to be a
monomial in the expansion of $Y$, gives the desired limit.  Summing the finitely many
monomials proves \eqref{eq:mixing-estimate} for every finite Wick polynomial $Y$.

Finally, finite Wick polynomials are dense in $L^2(\ma,\tau)$, and the approximation is
uniform in $\alpha$.  Indeed, if $Y_m$ is a finite Wick polynomial, then
\be
\begin{split}
\|(Y-Y_m)u_\alpha^{(I)}|\Omega\ra\|^2
&=
\tau\!\left(
\bigl(u_\alpha^{(I)}\bigr)^{\da}
(Y-Y_m)^{\da}(Y-Y_m)u_\alpha^{(I)}
\right) \\
&=
\|Y-Y_m\|_2^2,
\end{split}
\ee
where traciality and unitarity were used in the last step.  Moreover,
\be
\|\Phi_R(\eta)u_\alpha^{(I)}|\Omega\ra\|
\leq\|\Phi_R(\eta)\|
\ee
uniformly in $\alpha$.  Cauchy--Schwarz therefore extends
\eqref{eq:mixing-estimate} to every $Y\in\ma$.

\paragraph{Every central element must be trivial.}
Let $Z\in\mathcal Z(\ma)$ and take a finite chord vector
$|\eta\ra\perp\mh_H$.  Using a self-adjoint unitary
$u_\alpha^{(H)}\in\mb_H$, commutativity of the left and right algebras, and centrality
of $Z$, we find
\be
\begin{split}
\la\eta|Z|\Omega\ra
&=\la\Phi_R(\eta)\Omega|Z|\Omega\ra \\
&=\la u_\alpha^{(H)}\Phi_R(\eta)u_\alpha^{(H)}\Omega|Z|\Omega\ra \\
&=\la\Phi_R(\eta)u_\alpha^{(H)}\Omega
\bigm|Z u_\alpha^{(H)}\Omega\ra.
\end{split}
\ee
The right-hand side tends to zero by \eqref{eq:mixing-estimate}.  Finite chord vectors
orthogonal to $\mh_H$ are dense in $\mh_H^\perp$, and therefore
\be
Z|\Omega\ra\in\mh_H.
\ee
Repeating the same argument with $\mb_M$ gives
\be
Z|\Omega\ra\in\mh_M.
\ee
The pure $H$- and pure $M$-chord subspaces are orthogonal away from the
vacuum, so
\be
\mh_H\cap\mh_M=\mathbb C|\Omega\ra.
\ee
Hence $Z|\Omega\ra=\lambda|\Omega\ra$ for some $\lambda\in\mathbb C$.  Since
$|\Omega\ra$ is separating for $\ma$, this implies
\be \label{eq:trivial-center}
Z=\lambda\mathbf 1,
\qquad
\mathcal Z(\ma)=\mathbb C\mathbf 1.
\ee
Thus $\ma$ is a factor.

\paragraph{Type of the double-scaled algebra.}
The faithful normal trace \eqref{eq:vacuum-trace}, normalized by
$\tau(\mathbf 1)=1$, shows that $\ma$ is a finite factor.  It cannot be a finite
Type~I factor: every finite Type~I factor is a matrix algebra, whereas $\ma$ contains
the diffuse abelian algebra $\operatorname{vN}(H_L)$.  We therefore conclude that
\be
\ma=\operatorname{vN}(H_L,M_L)
\ee
is a Type~II$_1$ factor.  Its unique normalized trace is
\be
\tau(A)=\la\Omega|A|\Omega\ra,
\qquad A\in\ma.
\ee
By the left--right conjugation $J\ma_LJ=\ma_R$, the same conclusion holds for the right
chord algebra.

\section{Exploring Various Limits of Double-Scaled SYK} \label{sec:limits}
In this section, we delve into various limits of the double-scaled SYK model, examining its connections to other theories in detail.
\subsection{Revisiting the Triple Scaling Limit and its Connection to JT Gravity} \label{sec:triple}
In this section we discuss the triple scaling limit of DSSYK and examine the limiting result of $0$- and $1$-particle wavefunctions.  The triple scaling limit is characterized by setting the parameter $q=e^{-\lambda}$ in \eqref{eq:inner-product2} to $1$, while maintaining a constant value for $\lambda n=l$.
Here, the variable $n$ is related to the total number of chords in a typical state, and we will provide specific details when addressing the triple scaling limit of an individual state.  In the following discussion, we briefly review the derivation of the emergent Liouville Hamiltonian within this limit observed in \cite{Lin:2022rbf},  and
extend the discussion to $1$-particle case. We establish a dictionary between explicit expression of wavefunctions in DSSYK and their corresponding triple scaling limit. In addition, we provide interpretation in the context of JT gravity.  
\paragraph{$0$-Particle Wavefunction}
The action of the Hamiltonian on the $0$-particle spectrum gives rise to the following recursion relation for states with a fixed chord number \cite{Lin:2022rbf}:
\be \label{eq:recursion-0}
\frac{2\cos\te}{\sqrt{1-q}}\psi_{n}\left(\te\right)=\sqrt{[n+1]_{q}}\psi_{n+1}\left(\te\right)+\sqrt{[n]_{q}}\psi_{n-1}\left(\te\right),
\ee
where $\psi_n(\te)=\la\te|n\ra_N$ is the normalized wavefunction defined in section~\ref{sec:pure-DSSYK}. 
 Now we take triple scaling limit of \eqref{eq:recursion-0} with
length and energy given by:
\be
\begin{split}
    \left(1-q\right)[n]_{q}&=1-e^{-\lambda n}=1-\lambda^{2}e^{-\tilde{l}},\\
    \cos\te(s)&=\cos\lambda s\simeq1-\frac{1}{2}\lambda^{2}s^{2}+O\left(\lambda^{4}\right),
\end{split}
\ee
where in presenting chord number, we defined the renormalized length
$\tilde{l}$, which is related to $l$ by subtracting a divergent
constant: $\tilde{l}=l+2\log\lambda=\lambda n+2\log\lambda$. We also zoom in to the
edge of the energy spectrum by introducing $\te(s)=\lambda s$, where $\lambda$ is a small parameter, and $s\geq0$ serves as a new parameter for energy.  We now introduce the ``bulk'' wavefunction by switching the
energy and position in the original wave function, and express it in terms of the new parameters. More concretely, we introduce $\Psi_{s}(\tilde{l})$ as 
\be
\Psi_{s}(\tilde{l}):=\psi_{n(\tilde{l})}(\te(s)).
\ee
Now let us reformulate \eqref{eq:recursion-0} in terms of $\Psi_s (\tilde{l})$. We find it an identity at leading order in $O(\lambda)$. At next leading order, it becomes 
\be \label{eq:Liouville-0}
\left(-\partial_{\tilde l}^{2}+e^{-\tilde l}\right)\Psi_{s}\left(\tilde l\right)=s^{2}\Psi_{s}\left(\tilde l\right),
\ee
which turns out to be the equation satisfied by an energy eigenstate in Liouville quantum mechanics with energy $E_{\text{Liouville}}(s)=s^2$. The explicit solution to \eqref{eq:Liouville-0} can be expressed in terms of Bessel function as
\be
\Psi_{s}\left(\tilde l\right)=2K_{2is}\left(2e^{-\tilde l/2}\right),
\ee
where we have deduced the normalization constant of $\Psi_s$ by taking triple scaling limit of the normalization condition of $\psi_{n}\left(\te\right)$:
\be
\int_{0}^{\pi}\left(\mu\left(\te\right)\dd\te\right)\psi_{n}\left(\te\right)\psi_{m}\left(\te\right)=\del_{nm}.
\ee
The measure $\mu\left(\te\right)$ is specified in \eqref{eq:measure-0}, and
can be represented in terms of $q$-Gamma function as
\be
\mu\left(\te\right)=\frac{\left(q;q\right)_{\infty}^{3}\left(1-q\right)^{2}}{2\pi}\frac{1}{\Gamma_{q}\left(\pm2i\te/\lambda\right)}.
\ee
We drop out the $\te$-independent diverging constant in front when
taking triple scaling limit, and the integral measure becomes \cite{Berkooz:2018jqr}
\be
\int_{0}^{\pi}\dd\te\mu\left(\te\right)=C\left(\lambda\right)\int_{0}^{\pi/\lambda}\frac{\dd s}{2\pi}\frac{1}{\Gamma_{q}\left(\pm2is\right)}\stackrel{\lambda\to0}{\longrightarrow}\int_{0}^{\infty}\frac{\dd s}{2\pi}\frac{1}{\Gamma\left(\pm2is\right)}.
\ee
This reproduces the density of states in pure JT gravity:
\be
\rho(s)=\frac{1}{2\pi\Gamma\left(\pm2is\right)}=\frac{s}{2\pi^{2}}\sinh(2\pi s),
\ee
and by identifying $\psi_{n}\left(\te\right)$ with $2K_{2is}\left(2e^{-\tilde l/2}\right)$,
As a result, the normalization condition yields the appropriate normalization condition in the triple scaling limit:
\be
\int_{0}^{\infty}\dd s\rho\left(s\right)\left(2K_{2is}\left(2e^{-\tilde{l}/2}\right)\right)\left(2K_{2is}\left(2e^{-\tilde{l}^{\prime}/2}\right)\right)=\delta\left(\tilde{l}-\tilde{l}^{\prime}\right).
\ee
As a cross check, we can examine the generating function:
\be
\frac{1}{\left(e^{\pm i\te};q\right)_{\infty}}=\sum_{n=0}^{\infty}\frac{H_{n}\left(\cos\te|q\right)}{\left(q;q\right)_{n}}.
\ee
Continuing with a similar strategy, one can show that this relation accurately reproduces the following identity in the triple scaling limit:
\be
\int_{-\infty}^{\infty}d\tilde{l}\left(2K_{2is}(2e^{-\tilde{l}/2})\right)=\Gamma\left(\pm is\right).
\ee
In conclusion, we establish the following correspondence table between expressions in DSSYK and their counterpart in the triple scaling limit:
\be\begin{split}\label{eq:correspondence}
\psi_{n}\left(\te\right) & \longleftrightarrow2K_{2is}\left(2e^{-\tilde{l}/2}\right),\\
\int_{0}^{\pi}\mu\left(\te\right)\dd\te & \longleftrightarrow\int_{0}^{\infty}\rho\left(s\right)\dd s,\\
\sum_{n=0}^{\infty}\frac{H_{n}\left(\cos\te|q\right)}{\left(q;q\right)_{n}} & \longleftrightarrow\int_{-\infty}^{\infty}\dd\tilde{l}\left(2K_{2is}\left(2e^{-\tilde{l}/2}\right)\right).
\end{split}\ee
\paragraph{Matrix Components of Matter Operator }
Let us now move on to discuss the matter operator
$\mathcal{O}$ with weight $\Del$. In chord formulation
of double-scaled SYK , two point matter insertions with weight
$\Del$ is simply rephrased as insertion of $q^{\Del\hat{n}}$ in the chord Hilbert space, where $\hat{n}$
is the number operator of Hamiltonian chords. Note that this does
not belong to $\ma_{0}$ defined in section \ref{sec:pure-DSSYK} because it cannot be represented
as a bounded function of $H_0$. This can be seen from the observation that $q^{\Del\hat{n}}$ is not diagonal in the energy basis.
Instead, its components in the energy basis is given by:
\be \label{eq:size-operator}
\left\langle \theta_{1}\left|q^{\Del\hat n}\right|\theta_{2}\right\rangle =\sum_{n=0}^{\infty}q^{n\Del}\psi_{n}\left(\te_{1}\right)\psi_{n}\left(\te_{2}\right)=\frac{\left(q^{2\Del};q\right)_{\infty}}{\left(q^{\Del}e^{\pm i\te_{1}\pm i\te_{2}};q\right)_{\infty}}.
\ee
In evaluating the formula, we have inserted the identity $\mathbf{1}_{\mathcal{H}_{0}}=\sum_{n=0}^{\infty}|n\ra_N{}_N\!\la n|$.
We can then study the triple scaling limit of the above expression. In the zero-particle scaling, $q^n=\lambda^2e^{-\tilde l}$, and therefore $q^{\Del n}=\lambda^{2\Del}e^{-\Del\tilde l}$.  After removing this universal power of $\lambda$, and combining the result with the dictionary \eqref{eq:correspondence}, \eqref{eq:size-operator} becomes:
\be
\int_{-\infty}^{\infty}\dd\tilde l\, e^{-\Del\tilde l}\left(2K_{2is_{1}}\left(2e^{-\tilde l/2}\right)\right)\left(2K_{2is_{2}}\left(2e^{-\tilde l/2}\right)\right)=\frac{\Gamma\left(\Delta\pm is_{1}\pm is_{2}\right)}{\Gamma(2\Delta)}.
\ee
This precisely matches the two-point function $|\la E_{1}|\mathcal{O}|E_{2}\ra|^{2}$
in energy basis at disk level of JT gravity, with identification $E_{1/2}=s_{1/2}^{2}$.
The same formula can be obtained from boundary particle formalism
after fixing the $SL\left(2,\mathbb{R}\right)$ gauge \cite{Yang:2018gdb}, illustrated as:
\begin{equation}
    |\la E_1 | \mathcal{O} | E_2 \ra|^2=\begin{tikzpicture}[baseline={([yshift=-0.1cm]current bounding box.center)},scale=1]
        \draw[thick] (0,0) circle [radius=2];
        \draw[thick,blue] (-2,0)--(2,0);
        \node at (-2,0) [circle,fill,blue,inner sep=1.2pt]{};
        \node at (2,0) [circle,fill,blue,inner sep=1.2pt]{};   
        \node at (0,2.3) {$E_1$};
        \node at (0,-2.3) {$E_2$};
        \node at (0,0.3) {$e^{-\Delta l}$};
    \end{tikzpicture}
\end{equation}
In the current context, the theory lacks gauge redundancies in its description. Observables are unambiguously defined through their action on the physical Hilbert space. The outcome of the triple scaling limit, applied to relationships among observables in DSSYK, transforms into the corresponding relationships among gauge-invariant quantities in JT gravity. We adopt this perspective as a guiding principle in our subsequent exploration of the limit for $1$-particle wavefunctions.

\paragraph{One-particle Wavefunction}
We now move on to study the triple scaling limit of $1$-particle wavefunctions. A typical $1$-particle state in this case is labeled as $|n_{L},n_{R}\ra$
with left and right chord number specified as $n_{L}$ and $n_{R}$.
In appendix \ref{app:full-solution}, we provide a comprehensive derivation of the $1$-particle energy spectrum. In this section, we leverage the outcomes obtained in the appendix and explore the triple scaling limit of them. 

We denote the wavefunction in energy eigenbasis as $\psi_{n_{L},n_{R}}\left(\te_{L},\te_{R}\right)=\la\te_{L},\te_{R}|n_{L},n_{R}\ra$,
where $|\te_{L},\te_{R}\ra$ labels an eigenstate of both the left
and right Hamiltonian, with corresponding eigenvalue $2\cos\te_{L}/\sqrt{1-q}$
or $2\cos\te_{R}/\sqrt{1-q}$. Therefore, the action of left and right Hamiltonian $H_{L/R}$ yields the following two recursion relations of $\psi_{n_L,n_R}$:
\be\begin{split} \label{eq:recursion-1}
\frac{2\cos\te_{L}}{\sqrt{1-q}}\psi_{n_{L},n_{R}} & =\sqrt{[n_{L}+1]_{q}}\psi_{n_{L}+1,n_{R}}+\sqrt{[n_{L}]_{q}}\psi_{n_{L}-1,n_{R}}+q^{\Del+n_{L}}\sqrt{[n_{R}]}_{q}\psi_{n_{L},n_{R}-1},\\
\frac{2\cos\te_{R}}{\sqrt{1-q}}\psi_{n_{L},n_{R}} & =\sqrt{[n_{R}+1]_{q}}\psi_{n_{L},n_{R}+1}+\sqrt{[n_{R}]_{q}}\psi_{n_{L},n_{R}-1}+q^{\Del+n_{R}}\sqrt{[n_{L}]}_{q}\psi_{n_{L}-1,n_{R}}.
\end{split}\ee
They are related to the bi-variate $q$-Hermite functions in \eqref{eq:rec-2} via 
\be
\psi_{n_{L},n_{R}}\left(\te_{1},\te_{2}\right)=H_{n_{L},n_{R}}\left(\cos\te_{L},\cos\te_{R}|q,q^{\Del}\right)/\sqrt{\left(q;q\right)_{n_L}\left(q;q\right)_{n_R}}.
\ee
Now we discuss the triple scaling limit of \eqref{eq:recursion-1}. As before, we introduce the new set of parameters that stay finite in the limit as
\be
\begin{split}
\left(1-q\right)[n_{L/R}]_{q}&=1-e^{-\lambda n_{L/R}}=1-\lambda e^{-\tilde{l}_{L/R}}, \\
\cos\te_{L/R}(s_{L/R})&=\cos\lambda s_{L/R}\simeq1-\frac{1}{2}\lambda^{2}s_{L/R}^{2}+O\left(\lambda^{4}\right).
\end{split}
\ee
where the renormalized left and right length is defined as $\tilde{l}_{L/R}=\lambda n_{L/R}+\log\lambda$.
We then introduce the one particle wavefunction in terms of the new
parameters by switching the energy and position in $\psi_{n,m}$ as:
\be
\Psi_{s_{1},s_{2}}^{\Del}(\tilde{l}_{L},\tilde{l}_{R})=\psi_{n_L(\tilde{l}_{L}),n_R (\tilde{l}_{R})}\left(\te_{1}(s_{1}),\te_{2}(s_{2})\right).
\ee
By keeping the first non-trivial order of the two sides in the recursion
equation \eqref{eq:recursion-1}, we find that $\Psi_{s_{1},s_{2}}^{\Del}$ satisfies the following equations:
\be\begin{split} \label{eq:diff-1}
\left(-\partial_{\mathrm{L}}^{2}+e^{-\tilde{l}_{L}}\left(\Del+\partial_{R}-\partial_{L}\right)+e^{-\tilde{\ell}_{\mathrm{L}}-\tilde{\ell}_{\mathrm{R}}}\right)\Psi_{s_{1},s_{2}}^{\Del}(\tilde{l}_{L},\tilde{l}_{R}) & =s_{1}^{2}\Psi_{s_{1},s_{2}}^{\Del}(\tilde{l}_{L},\tilde{l}_{R}),\\
\left(-\partial_{\mathrm{R}}^{2}+e^{-\tilde{l}_{R}}\left(\Del-\partial_{R}+\partial_{L}\right)+e^{-\tilde{\ell}_{\mathrm{L}}-\tilde{\ell}_{\mathrm{R}}}\right)\Psi_{s_{1},s_{2}}^{\Del}(\tilde{l}_{L},\tilde{l}_{R}) & =s_{2}^{2}\Psi_{s_{1},s_{2}}^{\Del}(\tilde{l}_{L},\tilde{l}_{R}),
\end{split}
\ee
where $\partial_{R/L}$ are derivatives with respect to $\tilde{l}_{R/L}$.
Note that in deriving the equation we only assumed that $\Del$ is
$O\left(1\right)$ in $\lambda$. Therefore, the wavefunction
$\Psi_{s_{1},s_{2}}^{\Del}$ is only valid 
in the probing limit, where the matter does not back-react to the background geometry. Specifically, the geodesic length $\tilde{l}_{L/R}$ remains independent of the value of $\Delta$, which is a continuous parameter that does not scale with $\lambda$.  This observation is further substantiated by noting that $\Delta$ does not contribute to the energy spectrum. The energy of the gravitational state is characterized by two continuous parameters, denoted as $(s_{1},s_{2})$, ranging from $0$ to infinity.
In the following discussion, we consider two limits of \eqref{eq:diff-1} where
$\Psi_{s_{1},s_{2}}^{\Del}$ reduces to the familiar cases. 

The first limit is $\Del=0$, where one expects recovery of $0$-particle wavefunction. This assertion is supported by observing that:
\be
\lim_{\Del\to0}\Psi_{s_{1},s_{2}}^{\Del}\left(\tilde{l}_{L},\tilde{l}_{R}\right)=2\del\left(s_{1}-s_{2}\right)K_{2is_{1}}\left(2e^{-\left(\tilde{l}_{L}+\tilde{l}_{R}\right)/2}\right),
\ee
which says the $1$-particle wavefunction becomes the $0$-particle wavefunction
with total length $\tilde{l}=\tilde{l}_{L}+\tilde{l}_{R}$, and the energy spectrum is supported in the diagonal set with equal left and right energy $H_{L}=H_{R}$.
This aligns with the observation in pure JT gravity. In this situation,
there are no local energy fluctuations in bulk, and the left and right
Hamiltonian are equal after imposing the constraints \cite{Penington:2023dql}. 

The validity of the above equation can be confirmed by examining the two limits in a reversed ordering.  Let us consider taking $\Del\to0$ limit first in DSSYK by taking $r_V\to1$ in \eqref{eq:dp-1} with equal left and right energy. This leads to the linearization formula for $q$-Hermite polynomials
\be
\frac{H_{n_{L}}(\cos\te\mid q)H_{n_{R}}(\cos\te\mid q)}{(q;q)_{n_{L}}(q;q)_{n_{R}}}=\sum_{k=0}^{\min(n_{L},n_{R})}\frac{H_{n_{L}+n_{R}-2k}(\cos\te\mid q)}{(q;q)_{n_{L}-k}(q;q)_{n_{R}-k}(q;q)_{k}}.
\ee
The triple scaling limit of the two sides results in the following identity
\be \label{eq:composition}
\begin{split}
  \left(2K_{2is}\left(2e^{-\tilde{l}_{1}/2}\right)\right)\left(2K_{2is}\left(2e^{-\tilde{l}_{2}/2}\right)\right)=\int_{-\infty}^{\infty}\dd\tilde{l}&\exp\left(-e^{\left(\tilde{l}-\tilde{l}_{1}\right)/2}-e^{\left(\tilde{l}-\tilde{l}_{2}\right)/2}-e^{-\tilde{l}/2}\right) \\
& \times 2K_{2is}\left(2e^{\left(2\tilde{l}-\tilde{l}_{1}-\tilde{l}_{2}\right)/2}\right).
\end{split}
\ee
This is equivalent to the equation (261) of \cite{Lin:2022zxd}. 
From the boundary particle perspective,  \eqref{eq:composition} can be interpreted as the composition law of propagators of the boundary particle. Here
we obtain it as a fundamental relation between the $\Del\to0$ limit
of $1$-particle state and $0$-particle state. 

Now we move on towards another solvable case where we require $n_{L}=n_{R}$
and keep $\Del$ as $O(1)$ when taking the triple scaling
limit. We have $s_{L}=s_{R}$ simultaneously because of the left/right
symmetry. We then denote the wavefunction in this case as $\Psi_{s}^{\Del}(\tilde{l})$
where the total length is defined as: $\tilde{l}=2\tilde{l}_{L/R}$.
We find $\Psi_{s}^{\Del}(\tilde{l})$ satisfies the following
equation:
\be \label{eq:diff-2}
\left(-\partial_{\tilde l}^{2}+\Del e^{-\tilde{l}}+e^{-2\tilde{l}}\right)\Psi_{s}^{\Del}(\tilde{l})=s^{2}\Psi_{s}^{\Del}(\tilde{l}).
\ee
This corresponds to a particle moving in the Morse
potential, and was obtained in \cite{Gao:2021uro}
by quantizing JT gravity with end of world brane boundary conditions.
In that context, $\Del$ is related to the radial derivative of dilaton
field in the end-of-world brane boundary. The solution to \eqref{eq:diff-2}
can be explicitly written in terms of Whittaker function as:
\be
\Psi_{s}^{\Del}(\tilde{l})=e^{\tilde{l}/2}W_{-\frac{\Del}{2},is}(e^{-\tilde{l}}).
\ee
The normalization condition for $\Psi^{\Delta}_{s}$ can also be obtained by taking triple scaling limit of its counterpart in DSSYK.  It was
observed in \cite{okuyama2023end} that the triple
scaling limit of big-continuous $q$-Hermite function $\psi_{n}^{\Del}\left(\te\right)$
defined by the following recursion:
\be
\frac{2\cos\theta}{\sqrt{1-q}}\psi_{n}^{\Del}(\theta)=\sqrt{[n+1]_{q}}\psi_{n+1}^{\Del}(\theta)+q^{n+\frac{1+\Del}{2}}\psi_{n}^{\Del}(\theta)+\sqrt{[n]_{q}}\psi_{n-1}^{\Del}(\theta),
\ee
leads to the same equation as \eqref{eq:diff-2}. They are orthogonal with respect to a $\Delta$ dependent measure as:
\be
\int_{0}^{\pi}\frac{d\theta}{2\pi}\mu^{\Del}\left(\te\right)\psi_{n}^{\Del}(\te)\psi_{m}^{\Del}(\te)=\delta_{n,m}\qq\mu^{\Del}\left(\te\right)=\frac{\left(e^{\pm2i\te},q;q\right)_{\infty}}{\left(q^{\frac{\Del+1}{2}}e^{\pm i\te};q\right)_{\infty}}.
\ee
Taking the triple scaling limit of the two sides, and
with help of the correspondence table \eqref{eq:correspondence}, we obtain 
\be
\int_{0}^{\infty}\dd s\rho^{\Del}\left(s\right)\Psi_{s}(\tilde{l})\Psi_{s}(\tilde{l}^{\pp})=\delta\left(\tilde{l}-\tilde{l^{\prime}}\right),\qq\rho^{\Del}\left(s\right)=\rho\left(s\right)\Gamma\left(\frac{\Del+1}{2}\pm is\right).
\ee
For state $\Psi^{\Delta}_{s_L,s_R}(\tilde{l}_L,\tilde{l}_R)$ with general configuration in energy and length, we expect a derivation for an analytic expression by taking triple scaling limit of \eqref{eq:dp-2}. By taking triple scaling limit of \eqref{eq:dp-1}, we expect the following relation to hold,
\be
\left(2K_{2is_{1}}\left(2e^{-\tilde{l}_{1}/2}\right)\right)\left(2K_{2is_{2}}\left(2e^{-\tilde{l}_{2}/2}\right)\right)=\int_{-\infty}^{\infty}\dd\tilde{l}I^{\Del}\left(\tilde{l},\tilde{l}_{1},\tilde{l}_{2}\right)\Psi_{s_{1},s_{2}}^{\Del}\left(\tilde{l}-\tilde{l}_{1},\tilde{l}-\tilde{l}_{2}\right),
\ee
which relates the $1$-particle wavefunction with a product of two $0$-particle wavefunctions, and reduces to \eqref{eq:composition} in the $\Del\to0$ limit. The current results suggest that a rich amount of relationships for gauge-invariant quantities in JT gravity,
particularly in the probing limit, can be derived from the triple
scaling limit of double-scaled SYK. The systematic exploration of
extracting triple scaling limits from double-scaled SYK wavefunctions
is deferred to future research.

\subsection{The $q\to1$ Limit and its Connection to Hilbert Space of Baby Universes} \label{sec:baby}
We study the $q\to1$ limit with $r,r_{V}$ fixed, and  show how the chord inner product might be connected to the one in the Hilbert space of baby universe in this section.

In this limit, we can solve the recursive definition of the inner
product explicitly, and represent the result of \eqref{eq:inner-product} as:
\be\begin{split} \label{eq:inner-product-q2one}
\la n_{0},\cdots,n_{l}|m_{0},\cdots,m_{l}\ra|_{q\to1} & =\sum_{\pi\in S_{l}}\widetilde{\sum_{k_{ij}}}r^{\iota\left(\pi\right)}r_{V}^{\text{tr}\left(\mathbb{D^{\pi}}\mathbb{K}\right)}\cdot\frac{\prod_{i=0}^{l}n_{i}!m_{i}!}{\prod_{i,j=0}k_{ij}!}\\
 & =\sum_{\pi\in S_{l}}r^{\iota\left(\pi\right)}\widetilde{\sum_{k_{ij}}}r_{V}^{\text{tr}\left(\mathbb{D^{\pi}}\mathbb{K}\right)}\cdot\prod_{i=0}^{l}\binom{n_{i}}{k_{i1},\cdots,k_{il}}\prod_{j=0}^{l}m_{j}!,
\end{split}\ee
where the sum $\widetilde{\sum}_{\{k_{ij}\}}$ is defined as summing
over matrices $\mathbb{K}_{ij}=k_{ij}$ with the following constraint:
\be
\sum_{j=0}^{l}k_{ij}=n_{i},\qq\sum_{i=0}^{l}k_{ij}=m_{j},\qq i,j=0,1,\cdots,l.
\ee
The distance matrix $\mathbb{D}_{ij}^{\pi}$ in the exponent of
$r_{V}$ is defined by 
\be \label{eq:distance-matrix}
\mathbb{D}_{ij}^{\pi}=|i-j|+2c_{\pi}\left(i\right)\del_{ij},\qq i,j=0,1,\cdots,l.
\ee
The intuition for the formula is as follows: $q\to1$ means that the
crossings among Hamiltonian chords do not give rise to any penalty factor, so the inner
product \eqref{eq:inner-product-q2one} really counts the amount of ways of reassigning an
ensemble of $\left(n_{0},\cdots,n_{l}\right)$ Hamiltonian chords
into another ensemble $\left(m_{0},\cdots,m_{l}\right)$. This leads
to the constraint sum over $k_{ij}$ and the product of combinatoric
factor. The matrix element $k_{ij}$ is the number of chords at the $i$-th site in the initial state that have evolved into the $j$-th site in the final state. However, note that when this evolution is implemented, the
Hamiltonian chords must intersect with matter chords with $|i-j|$
times, which is the distance between the two sites and appears in
the distance matrix. The second term in \eqref{eq:distance-matrix} counts the extra
crossings due to the intersecting configuration of matter chords. 

Note that the above mechanism that determines the inner product is reminiscent of the baby universe Hilbert space developed in \cite{Giddings:1988cx} and later refined in \cite{Marolf:2020xie}. In that context, when computing the amplitude for evolving $n_i$ initial baby universes into $n_f$ final ones, one sums over all interpolating geometries. The dynamics allow any number $m$ of the initial universes to propagate into $m$ of the final ones, while additional baby universes may split off and later rejoin. 

The inner product of the baby universe Hilbert space is defined by summing over such geometries with weights determined by their genus. The process of evolving baby universes from an initial to a final configuration is structurally similar to connecting chords between initial and final chord states and summing over all possible configurations. The genus suppression associated with adding a handle to the geometry plays a role analogous to the penalty factor for an $H$–chord crossing a heavy $M$–chord. Because of these parallel structures, we expect a mathematical coincidence between the inner product \eqref{eq:inner-product-q2one} and that of the baby–universe Hilbert space. Below, we illustrate this correspondence explicitly in the one-particle sector.

The inner product \eqref{eq:inner-product-q2one} in one-particle sector reads:
\be
\frac{\left\langle n_{0},n_{1}\mid m_{0},m_{1}\right\rangle }{n_{0}!m_{0}!m_{1}!n_{1}!}
=\delta_{n_{0}+n_{1},m_{0}+m_{1}}\mi(n_{0},n_{1},m_{0})~,
\ee
where
\begin{equation} \label{I-def1}
\mi(n_{0},n_{1},m_{0})
=\sum_{k=0}^{\min\left(n_{1},m_{0}\right)}
\frac{r_{V}^{n_{0}-m_{0}+2k}}
{k!\left(n_{0}-m_{0}+k\right)!\left(n_{1}-k\right)!\left(m_{0}-k\right)!}~.
\end{equation}
We focus on the case with equal total chord numbers in the bra and ket states, omitting the Kronecker delta for brevity.  
For later convenience, we introduce an auxiliary variable $l$ and rewrite $\mathcal{I}$ as 
\be
\mi(n_{0},n_{1},m_{0})=\sum_{l=n_{0}-m_{0}}^{\infty}\sum_{k=0}^{{\rm min}(n_{1},m_{0})}\delta_{l,n_{0}-m_{0}+k}\frac{r_{V}^{k+l}}{k!l!(n_{1}-k)!(m_{0}-k)!}~,
\ee 
where for any $k\geq 0$, the sum over $l$ with the delta function sets $l=n_0-m_0+k\geq n_0 -m_0$, which reduces to the original expression \eqref{I-def1}.   Now consider summing over $n_0-m_0$ for fixed $m_0$:
\be
\sum_{n_{0}-m_{0}=-{\rm min}(n_{1},m_{0})}^{\infty}\frac{\left\langle n_{0},n_{1}\mid m_{0},m_{1}\right\rangle }{n_{0}!n_{1}!m_{0}!m_{1}!}=\sum_{n_{0}-m_{0}=-{\rm min}(n_{1},m_{0})}^{\infty}\mi(n_{0},n_{1},m_{0})~.
\ee
We can interchange the order of summation using
\be
\sum_{n_{0}-m_{0}=-{\rm min}(n_{1},m_{0})}^{\infty}
\sum_{l=n_{0}-m_{0}}^{\infty}
=\sum_{l=-{\rm min}(n_{1},m_{0})}^{\infty}
\sum_{n_{0}-m_{0}=-{\rm min}(n_{1},m_{0})}^{l}~,
\ee
and for fixed $m_0$, the sum over $n_0-m_0$ gives
\be
\sum_{n_{0}-m_{0}=-{\rm min}(n_{1},m_{0})}^{l}
\delta_{l,n_{0}-m_{0}+k}\left(\cdots\right)
=\Theta(-{\rm min}(n_{1},m_{0})\leq l-k\leq l)\left(\cdots\right)~,
\ee
where $\Theta$ is the step function (equal to $1$ if the condition is satisfied and $0$ otherwise).  
Thus we obtain:
\be\label{eq:inter-1}
\sum_{k=0}^{{\rm min}(n_{1},m_{0})}
\sum_{l=-{\rm min}(n_{1},m_{0})}^{\infty}
\Theta(-{\rm min}(n_{1},m_{0})\leq l-k\leq l)
\frac{r_{V}^{k+l}}{k!l!(n_{1}-k)!(m_{0}-k)!}~.
\ee
Since for $k\geq0$ the condition $l-k\leq l$ is always satisfied, we are left with $l\geq k-\text{min}(m_0,n_1)$. Hence \eqref{eq:inter-1} becomes
\be
\sum_{k=0}^{{\rm min}(n_{1},m_{0})}
\sum_{l=k-{\rm min}(n_{1},m_{0})}^{\infty}
\frac{r_{V}^{k+l}}
{k!l!(n_{1}-k)!(m_{0}-k)!}~.
\ee 
Because the integrand vanishes for negative $l$, and $k-\text{min}(n_1,m_0)\leq 0$, the lowest possible value of $l$ is always nonpositive. We can thus restrict to $l\ge0$, and the remaining sum over $l$ yields the exponential factor $e^{r_V}$. Therefore,
\be \label{eq:baby-resummation}
\sum_{d=-K}^{\infty}
\frac{\la m_0+d,n_1|m_0,n_1+d\ra}
{(m_0+d)!\,n_1!\,m_0!\,(n_1+d)!}
=e^{r_V}\sum_{k=0}^{K}
\frac{r_V^k}{k!\,(n_1-k)!\,(m_0-k)!},
\qquad K:=\min(n_1,m_0).
\ee

The right-hand side has the same splitting-and-rejoining combinatorics as equation (3.6) of~\cite{Giddings:1988cx}, up to an overall normalization and the identification
$r_V^{-1}=e^{-2S_0}(VT)^2$.  We regard this as a structural comparison rather than an identification of the full Hilbert spaces.  Here the baby universe Hamiltonian $H_{\text{BU}}$ is the operator defined in equation~(3.7) of~\cite{Giddings:1988cx}.

Based on the observations,  it is tempting to view that the limit considered in this section of
DSSYK might serve as certain completion of the Baby universe model. Practically, the combinatorial sum in~\eqref{eq:inner-product-q2one} is difficult to evaluate explicitly. In the following we present
an alternative formulation of \eqref{eq:inner-product-q2one}.  This is done by imposing the row constraints with Fourier phases and then carrying out the column-wise multinomial sums.  One obtains the exact representation
\be \label{eq:alternative-q2one}
\begin{split}
\left.\la n_0,\ldots,n_l|m_0,\ldots,m_l\ra\right|_{q\to1}
={}&\sum_{\pi\in S_l}r^{\iota(\pi)}
\left(\prod_{i=0}^l n_i!\right)
\int_0^{2\pi}\prod_{i=0}^l\frac{\dd\phi_i}{2\pi}\,
 e^{i\sum_i\phi_i n_i}\\
&\times\prod_{j=0}^l
\left(\sum_{i=0}^l e^{-i\phi_i}r_V^{\mathbb D_{ij}^{\pi}}\right)^{m_j}.
\end{split}
\ee
When $r_V=e^{-\lambda\Delta}$, the weight in the last line is
$e^{-i\phi_i-\lambda\Delta\mathbb D_{ij}^{\pi}}$.
The integral form of \eqref{eq:alternative-q2one} allows a saddle point approximation in evaluating the integral, extending its utility beyond the conventional regime where \(\Delta \sim O(1)\) in the \(\lambda \to 0\) limit. It would be interesting to explore the saddle point of \eqref{eq:alternative-q2one} while keeping the product \(\lambda \Delta\) large as \(\lambda \to 0\), which probes the semi-classical regime with heavy particle insertions.

\subsection{The $q\to0$ limit and its Connection to Brownian DSSYK}\label{sec:BDSSYK}
The authors in \cite{Milekhin:2023bjv} studied the algebra of Brownian DSSYK (BDSSYK) by explicitly constructing the algebra from the combined rules of
chord statistics in BDSSYK and Schwinger-Keldysh path integral. A typical feature of the chord rules is that Hamiltonian chords are prohibited from intersecting. This implies considering the limit $q\to 0$ when defining the inner product.   In this section we study the $q\to0$ limit with $r,r_V$ fixed of the algebra and states in section \ref{sec:DSSYK-matter}, and comment on its relation to BDSSYK. 

\paragraph{The $q\to0$ limit of $\mh_0$}
Let us first look at the $q\to0$ limit in the $0$-particle sector. This is a limit where crossings among Hamiltonian chords are not allowed. Therefore, it is easy to deduce that
\be
\la n |m\ra =\delta_{mn},
\ee
since for given amount of Hamiltonian chords, there is only one chord diagram that survives the limit. This is consistent with the fact that $\lim_{q\to0}[n]_q =1, \forall n\in\mathbb{Z}_{>0}$. The action of $H_0$ in this case yields the following recursion relation
\be
2\cos\te \psi_n(\theta)=\psi_{n+1}(\theta)+\psi_{n-1}(\theta),\quad \psi_{0}(\theta)=1,\psi_{-1}(\theta)=0,
\ee 
where we have used the fact that $[n]_{q=0}=1$. The solution to this recursion is given by Chebyshev polynomials of the second kind:
\be
\psi_n (\theta)= U_n (\cos\te)=\la\te|n\ra.
\ee
They are orthogonal under the Wigner measure:
\be
\int^{\pi}_0 \mu_0(\te) \dd \te \psi_n (\te) \psi_m (\te)=\del_{nm},\quad\mu_0(\te)=\frac{2}{\pi}\sin^{2}\te,
\ee
and we conclude that $\mh_0$ can be viewed as $L^2$-integrable functions in $[0,\pi]$ with this measure:
\be
\mathcal{H}_0=L^2([0, \pi], \mu_0(\theta)).
\ee
\paragraph{The $q\to0$ limit of $\mh$} 
We now consider taking $q\to0$ limit with $r,r_V$ fixed of the inner product \eqref{eq:inner-product}. In this limit, crossings between Hamiltonian chords are forbidden, and we only need to sum over configurations that involves Hamiltonian-matter crossings and matter-matter crossings. A typical chord diagram is depicted as follows:
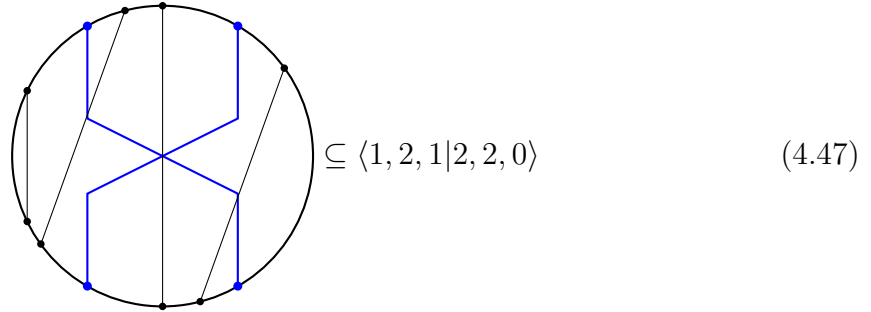
\begin{equation}
    \begin{tikzpicture}[baseline={([yshift=-0.1cm]current bounding box.center)},scale=1]
        \draw[thick] (0,0) circle [radius=2];
        \draw[thick,blue] (1,1.73) .. controls (0.5,0.6) and (-0.5,-0.) .. (-1,-1.73);
        \draw[thick,blue] (-1,1.73) .. controls (-.5,0.2) and (.5,-0.5) .. (1,-1.73);
        \draw plot [smooth] coordinates {(-1.618,-1.17)(-0.5,1.936)};
        \draw (0,2)--(0,-2);
        \draw (0.5,-1.936)--(1.618,1.17);
        \draw (-1.8,0.87)--(-1.8,-0.87);
        
        \node at (1,1.73) [circle,fill,blue,inner sep=1.2pt]{};
        \node at (-1,-1.73) [circle,fill,blue,inner sep=1.2pt]{};
        \node at (-1,1.73) [circle,fill,blue,inner sep=1.2pt]{};
        \node at (1,-1.73) [circle,fill,blue,inner sep=1.2pt]{};
        \node at (-1.618,-1.17) [circle,fill,black,inner sep=1pt]{};
        \node at (-0.5,1.936) [circle,fill,black,inner sep=1pt]{};
        \node at (0,2) [circle,fill,black,inner sep=1pt]{};
        \node at (0,-2) [circle,fill,black,inner sep=1pt]{};
        \node at (-1.8,0.87) [circle,fill,black,inner sep=1pt]{};
        \node at (-1.8,-0.87) [circle,fill,black,inner sep=1pt]{};
        \node at (1.618,1.17) [circle,fill,black,inner sep=1pt]{};
        \node at (0.5,-1.936) [circle,fill,black,inner sep=1pt]{};
    \end{tikzpicture} 
    \subseteq \la 1,2,1| 2,2,0\ra 
\end{equation}
We present the resulting inner product as follows:
\be \label{eq:inner-product-Brownian}
\la i_{0},\cdots,i_{k}|j_{0},\cdots j_{l}\ra=\del_{kl}\sum_{\pi\in S_{k}}r^{\iota\left(\pi\right)}r_{V}^{d_{0}\left(I,J\right)+d_{\pi}\left(I,J\right)}.
\ee
where $d_{0}\left(I,J\right)$ and $d_{\pi}\left(I,J\right)$ are
two discrete metric on the space of $k+1$-partitions $I=\{i_{0},\cdots,i_{k}\}$
and $J=\{j_{0},\cdots,j_{k}\}$ of integer $n=i_{0}+\cdots i_{k}=j_0+\cdots +j_l$.
They are defined as
\be\begin{split} \label{eq:discrete-metric}
d_{0}\left(I,J\right) & =\sum_{m=0}^{k}|\left(i_{0}-j_{0}\right)+\left(i_{1}-j_{1}\right)+\cdots+\left(i_{m}-j_{m}\right)|,\\
d_{\pi}\left(I,J\right) & =2\sum_{m=0}^{k}c_{\pi}\left(m\right)
\left[\min\{i_m,j_m\}-\left|\sum_{n=0}^{m-1}\left(i_n-j_n\right)\right|\right]_{+},
\end{split}\ee
Here $[x]_+:=\max\{x,0\}$. The sum $d_0+d_\pi$ correctly counts the total amount
of intersections between Hamiltonian chords and matter chords. $d_0$ counts
intersections arising from the evolution of Hamiltonian chords from one site in
the initial state to a distinct site in the final state. For $k$-th matter chord,
the amount of such intersections are given by the difference between
$i_0+\cdots+i_{k-1}$ and $j_0+\cdots +j_{k-1}$, which can be understood as
Hamiltonian chords that leak from the left to the right of the $k$-th matter
chord. Therefore, summing over $k$ counts the total number of intersections of
this particular kind. The other term $d_\pi$ counts the amount of Hamiltonian
chords that remain at the same site in the evolution. They intersects with matter
chords due to the intricate arrangement of matter chords. it is then easy to
deduce that the amount of those crossings are given by the second equation of
\eqref{eq:discrete-metric}, with explicit dependence on matter configuration
determined by $\pi$. 
\paragraph{connection to Brownian double-scaled SYK}
We conclude this subsection by pointing out potential connections to the Brownian model introduced in \cite{Milekhin:2023bjv}. Clearly, the fact that $q\to0$ models the situation where none of Hamiltonian chords can intersect among themselves. However, the Brownian model could live in a different representation of the algebra. To illustrate, let us consider the  $0$-particle sector. Note that the above limit of DSSYK yields a Hilbert space $\mh_0|_{q=0}$ with infinite dimension. However, in BDSSYK the Hilbert space $\mh^{{B}}_0$ associated with a single timefold without matter insertion is 1 dimensional, as one can collapse any amount of Hamiltonian chords to the empty state without creating any physical significance:
\begin{equation} \label{eq:def-Omega}
\begin{split}
        |\Omega\ra=\begin{tikzpicture}[baseline={([yshift=-0.1cm]current bounding box.center)},scale=1.5]
    \draw[thick] (0,0) -- (0.2,0) arc (90: -90: .15) -- (0,-.3) ;
    \end{tikzpicture} , \quad
    H_0 |\Omega\ra =\begin{tikzpicture}[baseline={([yshift=-0.1cm]current bounding box.center)},scale=1.5]
    \draw[thick] (0,0) -- (0.2,0) arc (90: -90: .15) -- (0,-.3) ;
    \draw[thick,black] (0.1,0) -- (0.1,-0.3);
    \end{tikzpicture}
    = |\Omega\ra .
\end{split}
\end{equation}
This difference can be understood as a choice of representation of the limiting relation $aa^\dagger=1$.  Its one-dimensional $*$-representations are
\be
a=e^{i\phi},\qquad a^\dagger=e^{-i\phi},\qquad H_0=2\cos\phi.
\ee
Thus a convention with $H_0|\Omega\ra=|\Omega\ra$ corresponds to $2\cos\phi=1$; it is a quotient representation and is distinct from the infinite-dimensional chord GNS representation $\mh_0|_{q=0}$. Yet by incorporating matter chord operators one can accommodate more states in the Hilbert space $\mh$ with matter, as studied in \cite{Milekhin:2023bjv} and \cite{Stanford:2023npy}. We expect similar form of inner product as in \eqref{eq:inner-product-Brownian} to appear in the context of Brownian DSSYK. The discussions presented in this paper are expected to be directly applicable in the context of Brownian DSSYK as well.

\section{Discussions and Future Prospective}\label{sec:future}
We conclude with future directions as follows:
\paragraph{Emergent temperature and Hyper-fast Scrambling}
We have shown that the empty state $\Omega$ satisfies a simplified version of KMS condition at infinite temperature, however, this does not mean that the theory
is insensitive to the finite temperature effect. Instead, the temperature
dependence is encoded in the operator algebra $\ma$. In \cite{Lin_2023} the authors proposed a geometrical realization of such dependence by incorporating it into coordinates
that parameterise the fake disk, which serves as a natural space for the symmetry algebra to act on. 

In our context, we can consider semi-classical limit of the operator
algebra and examine the correlation functions. As an example, let us
consider the operator $\Phi\left(n_{L},n_{R}\right)$ with 
\be
l=\lambda\left(n_{L}+n_{R}\right)=\lambda n=-2\log c,
\ee
where $c$ is fixed to be a constant when we take $\lambda\to0$ and
is related to the inverse temperature $\beta$ as $c=\cos\pi v/2=\pi v/\beta$.
Now we consider the two point function of operators $\Phi\left(n_{L},n_{R}\right)$
and $\Phi\left(n_{L}^{\pp},n_{R}^{\pp}\right)$ in this limit. Note
that the two-point function is vanishing unless $n_{L}+n_{R}=n_{L}^{\pp}+n_{R}^{\pp}$,
therefore, we introduce
\be
x=\frac{\lambda}{2}\left(n_{L}-n_{R}\right),\quad x^{\pp}=\frac{\lambda}{2}\left(n_{L}^{\pp}-n_{R}^{\pp}\right),
\ee
together with the following operators in the semi-classical limit
as: 
\be
\vp_{\beta}\left(x\right):=\lim_{\lambda\to0}\left[\frac{\lambda}{-2\log c}\right]^{1/2}\Phi\left(x-\frac{\log c}{\lambda},-x-\frac{\log c}{\lambda}\right),
\ee
where the $\beta$ dependence in $\vp_{\beta}\left(x\right)$ comes
from $c$. Then the result of two point function can be expressed
as 
\be
\la\Omega|\vp_{\beta}\left(x\right)\vp_{\beta}\left(x^{\pp}\right)|\Omega\ra=\left[\frac{\left(1-c^{2}\right)/2}{\cosh\frac{x-x^{\pp}}{2}-c\cosh\frac{x+x^{\pp}}{2}}\right]^{2\Del},
\ee
which exhibits explicit dependence on $c$, even though
$|\Omega\ra$ is an infinite temperature state with respect of the
operator algebra. This aligns with the observation in \cite{Lin:2022nss},\cite{Rahman:2024vyg} that a finite effective temperature can emerge and characterizes the thermal behavior of the system. In particular, the scrambling time depends on this emergent temperature instead of $\log N$, which is referred to as hyperfast and is conjectured to be a key feature of a putative holographic description of de Sitter gravity \cite{Susskind:2022bia},\cite{Susskind:2022dfz},\cite{Susskind:2023hnj}. 
it is natural to extend the above results to the entire
algebra $\ma$. Another natural question is whether the current discussion can be generalized to systems that exhibit similar emergence of temperature behavior.  It would be helpful to formulate an algebraic formalism that characterizes emergent effective temperature and hyperfast scrambling. 

\paragraph{Hagedorn regime and a possible Type~III$_1$ limit}
There is an alternative semi-classical regime of DSSYK that one expects
the algebra of one-sided operators to be of Type III$_{1}$. This
is the regime where one fixes $\beta\mathcal{J}$ and let $\lambda=2p^{2}/N$
goes to $0$. This is the regime where the collective field analysis
applies \cite{Goel:2023svz} and the
chord statics can be correctly reproduced by Liouville field theory
on a compactified causal wedge \cite{Lin_2023}.  In this
case, the partition function is given by
\begin{align*}
Z\left(\lambda\right) & =\int Dg\left(\tau_{1},\tau_{2}\right)e^{-I\left(\lambda\right)},\\
I\left(\lambda\right) & =\frac{\beta\mathcal{J}}{\lambda}\int_{[0,1]^{2}/\mathbb{Z}_{2}}\dd^{2}\tau\left(\partial_{\tau_{1}}g\partial_{\tau_{2}}g-e^{g\left(\tau_{1},\tau_{2}\right)}\right),
\end{align*}
which resembles the phase above the Hagedorn temperature
described in \cite{Witten:2021jzq,Witten:2021unn}, with $Z|_{\lambda\to0}=\infty$.  This divergence alone does not determine the von~Neumann type: a Type~III conclusion would require a specified limiting algebra and state, together with an analysis of their modular data.  It was pointed out in
\cite{leutheusser2023causal},\cite{leutheusser2023emergent} that the transition to Type III$_{1}$ can be probed
by the real two point function of single trace operators and the recent
paper \cite{gesteau2024explicit} raises a class of theories that exhibit such
transitions\footnote{In that context, however, the transition happens from Type I$_\infty$ to Type III$_1$. See also \cite{Liu:2024cmv} which suggests that distinct types of von Neumann algebra emerge, each accounting for different phases within a model of Majorana chains.}. It would be interesting to understand the phase structure
of double-scaled SYK and characterizes the transition of
the algebra constructed in the current work. 
\paragraph{The Switched Role of Energy and Position}
To obtain the bulk wavefunction in JT gravity, we have switched the
position and energy in the wavefunction of double-scaled SYK. We want
to point out that this is not by accident but a generic feature of
emergent gravitational interpretation. The fact that one needs to switch energy and position
to extract gravitational physics was also observed in the double scaling
limit of matrix models \cite{Saad:2019lba},\cite{Johnson:2021tnl}.  In the double-scaling limit, orthogonal polynomials in the matrix theory become continuous functions, and
by zooming in on the edge of the string equation simultaneously, one
obtains a dual quantum mechanical system. The energy and position
operators of this system are obtained by carefully taking the leading
order fluctuating part of the position and energy in the limiting
recursion relations of orthogonal polynomials. For a detailed derivation,
We refer the readers to equations (42)-(45) in \cite{Johnson:2022wsr}.

\paragraph{Entropy and Emergent Dilaton Profile}
A straightforward application of statements in \cite{Chandrasekaran:2022cip},\cite{Chandrasekaran:2022eqq} and \cite{Penington:2023dql}
shows that there is a notion of entropy for the matter chords algebra
unique up to constant rescaling. However, it is not clear how such an algebraic entropy would agree with the result in JT gravity in the semi-classical limit.
Specifically, a clear understanding of the emergence of a dilaton profile from the algebra $\ma$ in this limit is not available yet. Investigating this aspect further is a goal we aim to pursue in future research.

\paragraph{Delayed Scrambling and Hierarchy of Chaos} 
It was found in \cite{Rahman:2022jsf} that at time scale $t_* \sim \beta_{\text{GH}}\log(S)$ the light propagating fields in static patch start to contribute significantly to the 2-point function of operators localized at the stretched horizon. A further study in \cite{Milekhin:2024vbb} shows that the fact the two point function remains large for a relatively long time signatures a delay in scrambling\footnote{I thank Leonard Susskind for pointing this out.}, which is conjectured to happen for singlets in (charged-)DSSYK$_\infty$. The vast majority of entropy-carrying degrees of freedom exhibits hyper-fast scrambling without ever escaping the stretched horizon. One natural question arises: can we establish an algebraic framework for de Sitter that distinguishes these two distinct scrambling behaviors? If such a formulation exists, what potential connections might it unveil regarding the Hierarchy of chaos in Von Neumann algebra recently revisited in \cite{Gesteau:2023rrx}?

\section{Acknowledgement}
I would like to express my gratitude to Xi~Dong for helpful comments and suggestions on the draft of this paper.
 I appreciate valuable insights from Leonard~Susskind. I express my gratitude to Ahmed Almheri and Henry Lin for generously sharing their insights and 
providing detailed explanations on various aspects of double-scaled SYK.  I am thankful for collaborations with Elliott~Gesteau, Steven~B.~Giddings, Clifford~Johnson, and Alexey~Milekhin on related works. I thank Yiming~Chen, Gary~Horowitz, Adam~Levine, Don~Marolf, 
Vladimir Narovlansky,  Kazumi~Okuyama,  Xiaoliang~Qi, Sean~Mcbride, Mykhaylo~Usatyuk, Herman~Verlinde, Adel~Rahman, Douglas~Stanford, Eva~Silverstein, Stephen~H.~Shenker, Haifeng~Tang, Wayne W. Weng, Cynthia Yan, Shunyu~Yao and Ying~Zhao for helpful discussions. I have gained valuable insights and knowledge from the ongoing KITP Program “What is String Theory? Weaving Perspectives Together”. I extend my gratitude to the coordinators for organizing this wonderful event.  I acknowledge the support of the U.S. Department of Energy, Office of Science, Office of High Energy Physics, under Award Number DE-SC0011702.

\appendix
\section{Classification of von~Neumann Factors} \label{app:review}
In this section, we review the basic notions of von~Neumann factors and their classification.  
For a physically motivated discussion, see~\cite{Witten_2018}, and for a detailed and modern exposition of the mathematical classification, we refer to~\cite{Sorce_2023review}.

A von~Neumann algebra $\mathcal{A}$ is a $*$-subalgebra of bounded operators $\mathcal{B}(\mathcal{H})$ on a Hilbert space $\mathcal{H}$ that is closed under Hermitian conjugation and weak operator topology. Its center is defined as
\be
Z(\mathcal{A}) = \{\, a \in \mathcal{A} \;|\; [a,b]=0,  \forall b \in \mathcal{A} \,\}.
\ee
A von~Neumann algebra is called a \emph{factor} if its center is trivial, i.e.\ $Z(\mathcal{A})=\mathbb{C}\,\mathbf 1$. 
Equivalently, a factor has no nontrivial central decomposition, or superselection sectors detected by central projections.  This statement should not be confused with the absence of every possible tensor-product decomposition.

The Murray--von~Neumann classification applies precisely to such factors and is based on the equivalence classes of projection operators under partial isometries and on the existence (or absence) of a faithful trace functional.

\paragraph{Type~I.}
A factor $\mathcal{A}$ is of Type~I if it admits minimal projections, i.e.\ rank-one projectors exist within the algebra.  
Type~I factors are isomorphic to algebras of the form $\mathcal{B}(\mathcal{H})$, the bounded operators on a Hilbert space.  
Finite-dimensional examples $\mathrm{Mat}_n(\mathbb{C})$ are Type~I$_n$, while $\mathcal{B}(\mathcal{H})$ for infinite-dimensional $\mathcal{H}$ is Type~I$_\infty$.

\paragraph{Type~II.}
Type~II factors contain no minimal projections but admit a faithful normal semifinite trace.   They are subdivided into:
\begin{itemize}
    \item \textbf{Type~II$_1$:} factors with a normalized finite trace, $\mathrm{Tr}(\mathbf{1})=1$;\footnote{This condition is used to fix the normalization of the trace.}
    \item \textbf{Type~II$_\infty$:} infinite amplifications of Type~II$_1$, with a faithful normal semifinite trace satisfying $\operatorname{Tr}(\mathbf{1})=\infty$.
\end{itemize}
Relative to a chosen trace, normal states on a Type~II algebra can be represented by noncommutative $L^1$ densities, which supports a corresponding notion of entropy.

\paragraph{Type~III.}
Type~III factors admit no nontrivial finite trace: every nonzero projection is equivalent to the identity.  
They typically arise in local quantum field theory, where local operator algebras associated with spacetime subregions belong to this class (most notably Type~III$_1$).

\medskip
\noindent

Physically, a Type~II$_1$ algebra admits a faithful finite trace, allowing one to define trace densities and entropies.  
Because it has no minimal projections, it has no normal pure states; pure states may still exist in the broader $C^*$-algebraic sense.   This distinguishes Type~II factors from Type~I, where pure states exist, and from Type~III, where even the notion of a density matrix ceases to be well-defined.

\section{Towards a Full Solution of Energy Spectrum of 0- and 1-Particle Sector} \label{app:full-solution}
In this section we present a full solution to the energy spectrum by first constructing the generating function of wavefunctions of fixed length state in energy basis in $0$- and $1$-particle sector of $\mh$. We then show that the inner product in \cite{Lin_2023} are reproduced by integrating over the energy basis. This can be viewed as an independent derivation compared to the original $q$-weighted random walk approach. 
\subsection{The generating function of wavefunctions}
We start with the $0$-particle case. The action of $H_0$ on state $\psi_{n}(\te_1)$ is:
\be
\frac{2\cos\te}{\sqrt{1-q}}\psi_n (\te) = \sqrt{[n+1]}\,\psi_{n+1}(\te) + \sqrt{[n]}\,\psi_{n-1}(\te).
\ee
We now introduce $x=\cos\te$ and $H_n(x|q)=\sqrt{(q;q)_n} \psi_n$, we find the above relation becomes the standard recursion of $q$-Hermite polynomials:
\be
2xH_{n}\left(x;q\right)=H_{n+1}\left(x;q\right)+\left(1-q^{n}\right)H_{n-1}\left(x;q\right),
\ee
with boundary condition that $H_{-1}\left(x;q\right)=0,H_{0}\left(x;q\right)=1$.
We present a detailed derivation for its generating function and the strategy aligns with the latter solution of $1$-particle case in later discussion. we
introduce the following generating function:
\be
F_{0}\begin{bmatrix}x\\
s
\end{bmatrix}=\sum_{n=0}^{\infty}\frac{H_{n}\left(x;q\right)s^{n}}{\left(q;q\right)_{n}}.
\ee
The above recursion relation can then be written as
\be
2xF_{0}\begin{bmatrix}x\\
s
\end{bmatrix}=\frac{1}{s}\left(F_{0}\begin{bmatrix}x\\
s
\end{bmatrix}-F_{0}\begin{bmatrix}x\\
qs
\end{bmatrix}\right)+sF_{0}\begin{bmatrix}x\\
s
\end{bmatrix},
\ee
which can be presented in the following way
\be
F_{0}\begin{bmatrix}x\\
s
\end{bmatrix}=\frac{1}{1-2xs+s^{2}}F_{0}\begin{bmatrix}x\\
qs
\end{bmatrix}.
\ee
This is a $q$-difference equation. If we introduce $x=\cos\te$,
we find the denominator factorizes into
\be
1-2xs+s^{2}=\left(1-se^{i\te}\right)\left(1-se^{-i\te}\right),
\ee
and we can use the above recursion for infinite time, yielding:
\be\begin{split}
F_{0}\begin{bmatrix}x\\
s
\end{bmatrix} & =\lim_{k\to\infty}\frac{1}{\prod_{j=0}^{k}\left(1-q^{j}se^{i\te}\right)\left(1-q^{j}se^{-i\te}\right)}F_{0}\begin{bmatrix}x\\
q^{k}s
\end{bmatrix}\\
 & =\frac{1}{\left(se^{\pm i\te};q\right)_{\infty}}F_{0}\begin{bmatrix}x\\
0
\end{bmatrix}=\frac{1}{\left(se^{\pm i\te};q\right)_{\infty}},\qq |q|<1.
\end{split}\ee
where we have used the fact that $F_{0}=1$ for $s=0$, and we keep $|q|<1$ so that the infinite $q$-Pochhammer symbol is finite. 

Now we move on and solve for the following recursion relation induced from the action of $H_{L}$ on $|m,n\ra$:
\be \label{eq:rec-2}
\begin{aligned}2xH_{m,n}(x,y;q,r_V) & =H_{m+1,n}(x,y;q,r_V)\\
 & +\left(1-q^{m}\right)H_{m-1,n}(x,y;q,r_V)\\
 & +q^{m}\left(1-q^{n}\right)r_V H_{m,n-1}(x,y;q,r_V),
\end{aligned}
\ee
with $H_{m,0}\left(x,y;q,r_V\right)=H_{m}\left(x;q\right)$ and $H_{0,n}\left(x,y;q,r_V\right)=H_{n}\left(y\right)$.
$H_{m,n}$ is symmetric under left-right exchange:
\be
H_{m,n}\left(x,y;q,r_V\right)=H_{n,m}\left(y,x;q,r_V\right).
\ee
Therefore, we only need to solve \eqref{eq:rec-2} with above boundary conditions,
and the result will satisfy the recursion for $y$ automatically.
In the following, we shall leave the $q,r_V$ dependence of $H_{m,n}$
implicit for simplicity. We introduce the following generating function
\be
F_{1}\begin{bmatrix}x & y\\
s & t
\end{bmatrix}:=\sum_{m,n=0}^{\infty}\frac{H_{m,n}(x,y)s^{m}t^{n}}{(q;q)_{m}(q;q)_{n}}.
\ee
Similar derivation shows that the recursion of $H_{m,n}$ translates into the following $q$-difference equation of $F_{1}$: 
\be
F_{1}\begin{bmatrix}x & y\\
s & t
\end{bmatrix}=\frac{1-r_V st}{1-2xs+s^{2}}F_{1}\begin{bmatrix}x & y\\
qs & t
\end{bmatrix}
\ee
keep using the above recursion for $|q|<1$ we end up with: 
\be
F_{1}\begin{bmatrix}x & y\\
s & t
\end{bmatrix}=\frac{\left(r_V st;q\right)_{\infty}}{\left(se^{\pm i\te};q\right)_{\infty}}F_{1}\begin{bmatrix}x & y\\
0 & t
\end{bmatrix}=\frac{\left(r_V st;q\right)_{\infty}}{\left(se^{\pm i\te},te^{\pm i\phi};q\right)_{\infty}},
\ee
where we have set $x=\cos\te$ and $y=\cos\phi$, and used the fact
that 
\be
F_{1}\begin{bmatrix}x & y\\
0 & t
\end{bmatrix}=\frac{1}{\left(te^{\pm i\phi};q\right)_{\infty}}
\ee
in presenting the final result. In conclusion, we have:\footnote{The same generating function has been found in an earlier literature \cite{casper2020bivariate}, however, as we shall see in the following discussion, the measure in the current context differs from that in the paper and correctly reproduces the chord inner product in \cite{Lin_2023}.} 
\be \label{eq:gen1}
\frac{\left(r_V st;q\right)_{\infty}}{\left(se^{\pm i\te},te^{\pm i\phi};q\right)_{\infty}}=\sum_{m,n=0}^{\infty}\frac{H_{m,n}\left(x,y\right)s^{m}t^{n}}{\left(q;q\right)_{m}\left(q;q\right)_{n}},
\ee
which plays an important role in the computation of inner product
in subsequent discussion. 
\subsection{Evaluation of Inner Product with Energy Basis}
We show how to compute the inner product between fixed chord number
states by inserting the energy eigenbasis. For $0$-particle state,
we show that 
\be \label{eq:integral-0}
\la n_{1}|n_{2}\ra=\int_{0}^{\pi}\dd\te\mu\left(\te\right)\la n_{1}|\te\ra\la\te|n_{2}\ra,
\ee
where the overlap with the unnormalized number state is 
\be
\varphi_{n_{1}}\left(\te\right)=\la n_{1}|\te\ra=\frac{H_{n_{1}}\left(x\right)}{\left(1-q\right)^{n_{1}/2}},\qq x=\cos\te,
\ee
and the 0-particle measure 
\be \label{eq:measure-0}
\mu\left(\te\right)=\left(2\pi\right)^{-1}\left(e^{\pm2i\te},q;q\right)_{\infty},
\ee 
is deduced from the Jacobian when one moves from chord number basis
to energy basis. 

To evaluate the integral \eqref{eq:integral-0}, we consider the integral of the
generating function as follows:
\be \label{eq:integrand-0}
\int_{0}^{\pi}\dd\te\mu\left(\te\right)\frac{1}{\left(se^{\pm i\te};q\right)_{\infty}}\cdot\frac{1}{\left(te^{\pm i\te};q\right)_{\infty}}=\int_{0}^{2\pi}\frac{\dd\te}{2\pi}\frac{\left(e^{\pm2i\te},q;q\right)_{\infty}}{\left(se^{\pm i\te},te^{\pm i\te};q\right)_{\infty}}.
\ee
One can evaluate this integral by the Askey-Wilson integral given by equation (3.1.2) in \cite{koekoek1996askeyscheme},  the result reads
\be
\int_{0}^{2\pi}\frac{\dd\te}{2\pi}\frac{\left(e^{\pm2i\te},q;q\right)_{\infty}}{\left(se^{\pm i\te},te^{\pm i\te};q\right)_{\infty}}=\frac{1}{\left(st;q\right)_{\infty}}=\sum_{n=0}^{\infty}\frac{s^{n}t^{n}}{\left(q;q\right)_{n}}.
\ee
On the other hand, we can expand the integrand in \eqref{eq:integrand-0} as a double
series in $s$ and $t$, and integrate term by term, which yields
\be
\int_{0}^{2\pi}\frac{\dd\te}{2\pi}\frac{\left(e^{\pm2i\te},q;q\right)_{\infty}}{\left(se^{\pm i\te},te^{\pm i\te};q\right)_{\infty}}=\sum_{n,m=0}^{\infty}\int_{0}^{2\pi}\mu\left(\te\right)\frac{H_{n}\left(x\right)H_{m}\left(x\right)}{\left(q;q\right)_{n}\left(q;q\right)_{m}}\dd\te .
\ee
Therefore, we know
\be
\int_{0}^{2\pi}\mu\left(\te\right)H_{n}\left(x\right)H_{m}\left(x\right) \dd \theta=\del_{nm}\left(q;q\right)_{m},
\ee
and the integral in \eqref{eq:integral-0} becomes
\be
\la n_{1}|n_{2}\ra=\del_{n_{1},n_{2}}[n_{1}]!,
\ee
where we have introduced the $q$-factorial
\be
[n]_q !:=\frac{\left(q;q\right)_{n}}{\left(1-q\right)^{n}}.
\ee
The result matches the one derived by recursively using the $q$-commutation
relation. We now move on to the one-particle case, and we show the
following equation holds:
\be \label{eq:integral-1}
\la n_{L},n_{R}|n_{L}^{\pp},n_{R}^{\pp}\ra=\int\prod_{i=1}^{2}\mu\left(\te_{i}\right)\dd\te_{i}|\la\te_{1}|\mo|\te_{2}\ra|^{2}\la n_{L},n_{R}|\te_{1},\te_{2}\ra\la\te_{1},\te_{2}|n_{L}^{\pp},n_{R}^{\pp}\ra,
\ee
where the matter matrix element and wavefunction is defined as \cite{Berkooz:2018jqr}
\be\begin{split}
|\la\te_{1}|\mo|\te_{2}\ra|^{2} & =\frac{\left(r^{2}_V;q\right)_{\infty}}{\left(r_V e^{\pm i\te_{1}\pm i\te_{2}};q\right)_{\infty}},\\
\la\te_{1},\te_{2}|n_{L},n_{R}\ra & =\frac{H_{n_{1},n_{2}}\left(x,y\right)}{\sqrt{\left(1-q\right)^{n_{1}+n_{2}}}},\qq x=\cos\te_{1},y=\cos\te_{2},
\end{split}\ee
where $r_V =q^{\Del_V}$, $\Del_V$ is the conformal weight of the matter operator $V$. In
the following discussion, we keep $r_V$ and $q$ as two independent
parameters that ranges from $0$ to $1$. Following the same strategy,
we evaluate \eqref{eq:integral-1} by considering the integral of the generating
function:
\be\begin{split} \label{eq:integrand-1}
I\left(s_{1},s_{2},t_{1},t_{2}\right) & :=\int_{0}^{\pi}\left(\prod_{i=1}^{2}\mu\left(\te_{i}\right)\dd\te_{i}\right)|\la\te_{1}|\mo|\te_{2}\ra|^{2}\cdot\frac{\left(r_V s_{1}t_{1},r_V s_{2}t_{2};q\right)_{\infty}}{\left(s_{1}e^{\pm i\te_{1}},s_{2}e^{\pm i\te_{1}},t_{1}e^{\pm i\te_{2}},t_{2}e^{\pm i\te_{2}};q\right)_{\infty}}\\
 & =\int_{[0,\pi]^{2}}\frac{\dd\te_{1}\dd\te_{2}}{\left(2\pi\right)^{2}}\frac{\left(r_V s_{1}t_{1},r_V s_{2}t_{2},r^{2}_V,e^{\pm2i\te_{1}},e^{\pm2i\te_{2}},q,q;q\right)_{\infty}}{\left(s_{1}e^{\pm i\te_{1}},s_{2}e^{\pm i\te_{1}},t_{1}e^{\pm i\te_{2}},t_{2}e^{\pm i\te_{2}},r_V e^{\pm i\te_{1}\pm i\te_{2}};q\right)_{\infty}}.
\end{split}\ee
We evaluate the $\te_{1}$ integral first by Askey-Wilson formula, we
find 
\be
\int_{0}^{\pi}\frac{\mathrm{d}\theta_{1}}{2\pi}\frac{\left(e^{\pm2i\te_{1}},q;q\right)_{\infty}}{\left(s_{1}e^{\pm i\te_{1}},s_{2}e^{\pm i\te_{1}},r_V e^{\pm i\te_1\pm i\te_2};q\right)_{\infty}}=\frac{\left(s_{1}s_{2}r^{2}_V;q\right)_{\infty}}{\left(s_{1}s_{2},r^{2}_V;q\right)_{\infty}\left(s_{1}r_V e^{\pm i\te_{2}},s_{2}r_V e^{\pm i\te_{2}};q\right)_{\infty}}.
\ee
The subsequent integral over $\te_{2}$ gives:
\be
\begin{split}
    \int_{0}^{\pi}&\frac{\dd\te_{2}}{2\pi}\frac{\left(e^{\pm2i\te_{2}},q;q\right)_{\infty}}{\left(s_{1}r_V e^{\pm i\te_{2}},s_{2}r_V e^{\pm i\te_{2}},t_{1}e^{\pm i\te_{2}}, t_{2}e^{\pm i\te_{2}};q\right)_{\infty}}\\
&=\frac{\left(s_{1}s_{2}t_{1}t_{2}r^{2}_V;q\right)_{\infty}}{\left(s_{1}s_{2}r^{2}_V,s_{1}t_{1}r_V,s_{1}t_{2}r_V,s_{2}t_{1}r_V,s_{2}t_{2}r_V,t_{1}t_{2};q\right)_{\infty}}.
\end{split}
\ee
Therefore, we know
\be
I\left(s_{1},s_{2},t_{1},t_{2}\right)=\frac{\left(s_{1}s_{2}t_{1}t_{2}r^{2}_V;q\right)_{\infty}}{\left(s_{1}s_{2},t_{1}t_{2},s_{1}t_{2}r_V,s_{2}t_{1}r_V;q\right)_{\infty}}.
\ee
We can expand the integrand of \eqref{eq:integrand-1} and exchange the integral and sum, which yields
\be
\begin{split}
I(s_{1},s_{2},t_{1},t_{2})=&\sum_{n_{L},n_{R},n_{L}^{\pp},n_{R}^{\pp}=0}^{\infty}\int_{0}^{\pi}\left(\prod_{i=1}^{2}\mu\left(\te_{i}\right)\dd\te_{i}\right)|\la\te_{1}|\mo|\te_{2}\ra|^{2} \\
&\times \frac{H_{n_{L},n_{R}}H_{n_{L}^{\pp},n_{R}^{\pp}}}{(q;q)_{n_L}(q;q)_{n_R}(q;q)_{n^{\pp}_L}(q;q)_{n^{\pp}_R}}s_{1}^{n_{L}}t_{1}^{n_{R}}s_{2}^{n_{L}^{\pp}}t_{2}^{n_{R}^{\pp}}.
\end{split}
\ee
by matching order by order in $\left(s_{1},t_{1},s_{2},t_{2}\right)$,
we conclude that for $n_{L}\geq n^{\pp}_{L}$, we have
\be \label{eq:inner-product-11}
\begin{split}
\int_{[0, \pi]^2}( & \prod_{i=1}^2 \mu\left(\theta_i\right) \mathrm{d} \theta_i\left|\left\langle\theta_1|\mathcal{O}| \theta_2\right\rangle\right|^2 H_{n_L, n_R} H_{n_L^{\prime}, n_R^{\prime}}=\delta_{n_L+n_R, n_L^{\prime}+n_R^{\prime}} \times \\
& \left(\sum_{k=0}^{\min \left(n_R, n^{\pp}_L\right)} q^{k^2+k\left(n_L-n_L^{\prime}\right)} r^{2 k+n_L-n^{\pp}_L}_V \frac{(q ; q)_{n_L}(q ; q)_{n_R}(q ; q)_{n_L^{\prime}}(q ; q)_{n_L+n_R-n_L^{\prime}}}{(q ; q)_{n_L^{\prime}-k}(q ; q)_{n_R-k}(q ; q)_{n_L-n_L^{\prime}+k}(q ; q)_k}\right).
\end{split}
\ee
it is straightforward to show
that the result matches \cite{Lin_2023} by converting $H_{n_L, n_R}$ to the corresponding normalized wavefunction $\psi_{n_L,n_R}$.

A general state with arbitrary amount of matter chords can be constructed from $0$- and $1$-particle states through block decomposition. For example, a general $2$-particle state $|n_L,n_1,n_R\ra$ can be decomposed into chord irreducible representations as \cite{Lin_2023}:
\be
\left|n_L, n_1, n_R\right\rangle=\sum_{m_L+m_R+k=n_1} \psi_{k, m_L, m_R}\left|[\mathrm{VW}]_k ; n_L+m_L, n_R+m_R\right\rangle,
\ee
with Clebsch-Gordon coefficients $\psi_{k,m_L,m_R}$. Therefore, we expect that the results present in this section to extend to the entire Hilbert space $\mh$ and determines the full energy spectrum of the theory. 

\section{The Fock-Decomposition of Lin-Stanford Basis} \label{app:Fock}
In this section we try to clarify the relationship between Lin-Stanford basis and Fock space basis, and point out a reformulation of the inner product in terms of a doubled Hilbert space, as suggested in \cite{Okuyama:2024yya}.   For clarity, in the following discussion we denote Lin-Stanford
basis with $k$ matter chords as $|n_{1},\cdots,n_{k}\ra$
and the Fock basis as $|n_{1}\ra\otimes\cdots\otimes|n_{k}\ra$. 
From the discussion in section \ref{sec:pure-DSSYK}, it is clear that the $0$-particle states
are equivalent. We then move on to discuss $1$-particle
states. 

We observe that the generating function of $1$-particle states
can be re-expressed in terms of $0$-particle states as:
\be \label{eq:gen2}
\frac{(r_V st;q)_{\infty}}{\left|\left(se^{i\theta},te^{i\phi};q\right)_{\infty}\right|^{2}}=\sum_{k,m,n=0}^{\infty}\frac{(-1)^{k}q^{\binom{k}{2}}r^{k}_V s^{m+k}t^{n+k}}{(q;q)_{k}(q;q)_{m}(q;q)_{n}}H_{m}(x)H_{n}(y).
\ee
Therefore, by comparing \eqref{eq:gen2} with \eqref{eq:gen1}, we conclude that the tensor product of two $0$-particle wavefunction can be expressed in terms of superposition of $1$-particle wavefunctions as:
\be \label{eq:dp-1}
\frac{H_{n}\left(x|q\right)H_{m}\left(y|q\right)}{\left(q;q\right)_{n}\left(q;q\right)_{m}}=\sum_{k=0}^{{\rm min}\left(m,n\right)}\frac{r^{k}_V H_{n+m-2k}\left(x,y|q,r_V\right)}{\left(q;q\right)_{n-k}\left(q;q\right)_{m-k}\left(q;q\right)_{k}}.
\ee
The inverse of the above formula then presents a way to express $H_{m,n}$ in terms of a weighted summation of product of $H_n$s:
\be \label{eq:dp-2}
H_{m, n}(x, y)=\sum_{k=0}^{\min (m, n)} \frac{(-1)^k q^{k(k-1)/2}(q ; q)_m(q ; q)_n r^{k}_V}{(q ; q)_{m-k}(q ; q)_{n-k}(q ; q)_k} H_{m-k}(x ; q) H_{n-k}(y ; q)
\ee
The formula can be understood as mapping $|m,n\ra$ to a Fock space of $\mh_0$, where the resulting state sums over all possible pairings between open chords in the left and right. Each crossing between Hamiltonian chords contributes to a weight of $q$ and each crossing between Hamiltonian chords and matter chords contributes to a weight of $-r_V$. The coefficient in \eqref{eq:dp-2} correctly counts the result of the weighted sum. This map can be illustrated as:\footnote{I thank Ahmed Almheiri for helpful discussions on this point.}
\begin{equation}
\begin{tikzpicture}[baseline={([yshift=-0.1cm]current bounding box.center)},scale=1]
    \draw[thick] (2,0) arc (0:-180: 2);
    \draw[thick,blue] (0,-2)--(0,0);
    \node at (0,-2) [circle,fill,blue,inner sep=1.2pt]{};
    \draw (0.364,-1.966)--(0.364,0);
    \draw (0.673,-1.883)--(0.673,0);
    \draw (1.605,-1.192)--(1.605,0);
    \draw (-0.37,-1.965)--(-.37,0);
    \draw (-1.78,-.901)--(-1.78,0);
    \draw (-1.265,-1.54)--(-1.265,0);
    \draw (-1.51,-1.31) -- (-1.51,0);
    \node at (0.364,-1.966) [circle,fill,black,inner sep=1pt]{};   
    \node at (0.673,-1.883) [circle,fill,black,inner sep=1pt]{};
    \node at (1.605,-1.192) [circle,fill,black,inner sep=1pt]{};
    \node at (-0.37,-1.965) [circle,fill,black,inner sep=1pt]{};
    \node at (-1.78,-.901) [circle,fill,black,inner sep=1pt]{};
    \node at (-1.265,-1.54) [circle,fill,black,inner sep=1pt]{};
    \node at (-1.51,-1.31) [circle,fill,black,inner sep=1pt]{};
    \node at (0,-2.3) {$|m,n\ra $};
\end{tikzpicture}\simeq \sum_{k=0}^{\min(m,n)}\sum_{\substack{\text{configurations} \\ \text {with $k$ crossings}}}
\begin{tikzpicture}[baseline={([yshift=-0.1cm]current bounding box.center)},scale=1]
    \draw[thick] (0,0) arc (0:-180: 1);
    \draw[thick] (0,0) arc (-180:0: 1);
    \draw (-1.807,-0.589)--(-1.807,0);
    \draw (-1.528,-0.848)--(-1.528,0);
    \draw (1.699,-0.714)--(1.699,0);
    \draw plot [smooth] coordinates {(-0.546,-0.891)(-0.2,0)(0,0.2)(0.2,0)(0.546,-0.891)};
    \draw plot [smooth] coordinates {(-1,-1)(-0.5,0)(0,0.5)(0.5,0)(1,-1)};
    \draw[thick,blue] (0,0)--(0,1);
    \node at (0,0) [circle,fill,blue,inner sep=1.2pt]{};
    \node at (-1.807,-0.589) [circle,fill,black,inner sep=1pt]{};
    \node at (-1.528,-0.848) [circle,fill,black,inner sep=1pt]{};
    \node at (1.699,-0.714) [circle,fill,black,inner sep=1pt]{};
    \node at (-0.546,-0.891) [circle,fill,black,inner sep=1pt]{};
    \node at (0.546,-0.891) [circle,fill,black,inner sep=1pt]{};
    \node at (-1,-1) [circle,fill,black,inner sep=1pt]{};
    \node at (1,-1) [circle,fill,black,inner sep=1pt]{};
    \node at (0,-1.3) {$|m-k\ra \otimes |n-k\ra $};
    \node at (1.2,0.6) {$k$-crossings};
\end{tikzpicture}.
\end{equation}
Now we extrapolate the above relations a little bit and try to formulate the inner product \eqref{eq:inner-product-11} in a doubled Hilbert space. We do this by rewriting \eqref{eq:dp-2} as:
\be\begin{split} \label{eq:relation}
|m,n\ra & =\sum_{k=0}^{{\rm min}\left(m,n\right)}\frac{(-1)^{k}q^{\binom{k}{2}}(q;q)_{m}(q;q)_{n}r^{k}_V}{(q;q)_{m-k}(q;q)_{n-k}(q;q)_{k}}|m-k\ra\otimes|n-k\ra\\
 & =|m\ra\otimes|n\ra+\left(\text{states with total chord number less than }m+n\right),
\end{split}\ee
from which one can deduce the linear-independence of Lin-Stanford basis. We explore further this relation and denote \eqref{eq:relation} systematically as
\be \label{eq:dp-4}
|m,n\ra=\sum_{k=0}^{{\rm min}\left(m,n\right)}c_{m,n}\left(k\right)|m-k\ra\otimes|n-k\ra,\qq c_{m,n}\left(0\right)=1.
\ee
The inner product discussed in the previous section can be expressed
as:
\be\begin{split}
\la m,n|m^{\pp},n^{\pp}\ra & =\int\prod_{i=1}^{2}\mu\left(\te_{i}\right)\dd\te_{i}|\la\te_{1}|\mo|\te_{2}\ra|^{2}\sum_{k=0}^{{\rm min}\left(m,n\right)}\sum_{k^{\pp}=0}^{{\rm min}\left(m^{\pp},n^{\pp}\right)}c_{m,n}\left(k\right)c_{m^{\pp},n^{\pp}}\left(k^{\pp}\right)\\
 & \times\left(\la m-k|\te_{1}\ra\la n-k|\te_{2}\ra\la\te_{1}|m^{\pp}-k^{\pp}\ra\la\te_{2}|n^{\pp}-k^{\pp}\ra\right).
\end{split}\ee
The matter density of state can be interpreted as a two point function
in the double Hilbert space as:
\be
|\la\te_{1}|\mo|\te_{2}\ra|^{2}=\la\te_{1},\te_{2}|\mo^{L}\mo^{R}|\te_{2},\te_{1}\ra,
\ee
where we have embedded the original operator $\mo$ as an 2-sided
operator as $\mo^{L}:=\mo\otimes1_{R}$ and similar for $\mo^{R}$.
Therefore, we find the integral becomes
\be
\begin{split}
\int\prod_{i=1}^{2}&\left(\mu\left(\te_{i}\right)\dd\te_{i}\right)
\times \\
& \left(\la m-k|\otimes\la n-k|\right)|\te_{1},\te_{2}\ra\la\te_{1},\te_{2}|\mo^{L}\mo^{R}|\te_{1},\te_{2}\ra\la\te_{1},\te_{2}|\left(|m^{\pp}-k^{\pp}\ra\otimes|n^{\pp}-k^{\pp}\ra\right).
\end{split}
\ee
The diagrammatic rules developed in \cite{Berkooz:2018jqr} suggest that 
\be
\la\te_{1},\te_{2}|\mo^{L}\mo^{R}|\te_{3},\te_{4}\ra\propto\mu^{-1}\left(\te_{3}\right)\mu^{-1}\left(\te_{4}\right)\del\left(\te_{3}-\te_{2}\right)\del\left(\te_{4}-\te_{1}\right).
\ee
Therefore, we can rewrite the above equation as 
\be
\begin{split}
\int\prod_{i=1}^{4}&\left(\mu\left(\te_{i}\right)\dd\te_{i}\right)\times\\
&\left(\la m-k|\otimes\la n-k|\right)|\te_{1},\te_{2}\ra\la\te_{1},\te_{2}|\mo^{L}\mo^{R}|\te_{3},\te_{4}\ra\la\te_{3},\te_{4}|\left(|m^{\pp}-k^{\pp}\ra\otimes|n^{\pp}-k^{\pp}\ra\right).
\end{split}
\ee
Now with the completeness relation, we know this is equivalent to
\be
(\la m-k|\otimes\la n-k|)\mo^{L}\mo^{R}(|m^{\pp}-k^\pp\ra\otimes|n^{\pp}-k^\pp\ra).
\ee
Combined with \eqref{eq:dp-4}, we conclude with
\be
\begin{split}
    \la m,n|m^{\pp},n^{\pp}\ra=\sum_{k=0}^{{\rm min}\left(m,n\right)}\sum_{k^{\pp}=0}^{\min\left(m^{\pp},n^{\pp}\right)}& c_{m,n}(k)c_{m^{\pp},n^{\pp}}(k^{\pp})\times \\
&\left(\la m-k|\otimes\la n-k|\mo^{L}\mo^{R}|m^{\pp}-k^{\pp}\ra\otimes|n^{\pp}-k^{\pp}\ra\right).
\end{split}
\ee
Therefore, we deduce that the inner product between Lin-Stanford $1$-particle states can be mapped to two point correlators in Fock basis. The resulting Fock states are obtained by summing over all possible pairings of open chords between the two sides in the original state.

\section{A Bound on the $Q$-Symmetrizer} \label{app:Bound-PQ}

We prove the following operator inequality on
$\mathfrak h^{\otimes(N+1)}$:
\begin{equation}
    P_Q^{(N+1)}
    \leq
    {1\over 1-\rho}\,
    \left(1\otimes P_Q^{(N)}\right),
    \label{eq:Q-symmetrizer-bound}
\end{equation}
where $\mathbf 1\otimes P_Q^{(N)}$ acts trivially on the first tensor
factor and as $P_Q^{(N)}$ on the remaining $N$ factors.  Here
\[
    \rho=\max_{I,J}|Q_{IJ}|<1 .
\]
Let
\begin{equation}
    T(e_I\otimes e_J)=Q_{IJ}\,e_J\otimes e_I ,
\end{equation}
and let $T_i$ denote the operator which acts as $T$ on the $i$-th and
$(i+1)$-st tensor factors.  Since $Q_{IJ}=Q_{JI}$, the operator $T$ is
self-adjoint.  Moreover, the operators $T_i$ satisfy the Yang-Baxter relations.
Thus, if
\[
    \sigma=s_{i_1}\cdots s_{i_\ell}
\]
is a reduced expression for a permutation $\sigma\in S_N$, the operator
\begin{equation}
    \varphi(\sigma)=T_{i_1}\cdots T_{i_\ell}
\end{equation}
is independent of the chosen reduced expression.  The $Q$-symmetrizer is
then
\begin{equation}
    P_Q^{(N)}=\sum_{\sigma\in S_N}\varphi(\sigma).
\end{equation}

We regard $S_N$ as the subgroup of $S_{N+1}$ which fixes the first tensor
factor and permutes the last $N$ tensor factors.  Equivalently, this
subgroup is generated by $s_2,\ldots,s_N$.  Every permutation
$\sigma\in S_{N+1}$ has a unique length-additive decomposition
\begin{equation}
    \sigma=(s_k s_{k-1}\cdots s_1)\tau ,
    \qquad
    0\leq k\leq N,
    \qquad
    \tau\in S_N ,
\end{equation}
where for $k=0$ the first factor is understood to be the identity.  The
integer $k$ records the final position, $k+1$, of the first tensor factor.
Length additivity gives
\begin{equation}
    \varphi(\sigma)
    =
    \varphi(s_k s_{k-1}\cdots s_1)\,\varphi(\tau).
\end{equation}
Therefore
\begin{equation}
\begin{aligned}
    P_Q^{(N+1)}
    &=
    \sum_{\sigma\in S_{N+1}}\varphi(\sigma)                                      \\
    &=
    \sum_{k=0}^N
    \varphi(s_k s_{k-1}\cdots s_1)
    \left(1\otimes P_Q^{(N)}\right)                                               \\
    &=
    \mathcal R_{N+1}\left(1\otimes P_Q^{(N)}\right),
\end{aligned}
\label{eq:PQ-recursion}
\end{equation}
where
\begin{equation}
    \mathcal R_{N+1}
    =
    1+T_1+T_2T_1+\cdots+T_NT_{N-1}\cdots T_1 .
\end{equation}
Since $\|T_i\|\leq \rho$, we have
\begin{equation}
    \|\mathcal R_{N+1}\|
    \leq
    \sum_{k=0}^N \rho^k
    \leq
    {1\over 1-\rho}.
    \label{eq:R-bound}
\end{equation}

It remains to convert this norm estimate into the desired quadratic-form
inequality.  Set
\begin{equation}
    A_N=1\otimes P_Q^{(N)} .
\end{equation}
For $\rho<1$, the symmetrizers are strictly positive on each fixed length
sector, so $A_N$ is positive and invertible.  Define
\begin{equation}
    B_N
    =
    A_N^{-1/2}P_Q^{(N+1)}A_N^{-1/2}.
\end{equation}
The operator $B_N$ is positive and self-adjoint.  Using
\eqref{eq:PQ-recursion}, we may also write
\begin{equation}
    B_N
    =
    A_N^{-1/2}\mathcal R_{N+1}A_N^{1/2}.
\end{equation}
Thus $B_N$ is similar to $\mathcal R_{N+1}$, and hence has the same
spectrum.  Since $B_N$ is positive,
\begin{equation}
    \|B_N\|
    \leq
    \|\mathcal R_{N+1}\|
    \leq
    {1\over 1-\rho},
\end{equation}
Therefore
\begin{equation}
    0\leq B_N\leq {1\over 1-\rho}\, \mathbf{1} .
\end{equation}
Multiplying by $A_N^{1/2}$ on the left and right gives
\begin{equation}
    P_Q^{(N+1)}
    \leq
    {1\over 1-\rho}\,
    A_N
    =
    {1\over 1-\rho}
    \left(1\otimes P_Q^{(N)}\right),
\end{equation}
which proves \eqref{eq:Q-symmetrizer-bound}.

\section{Detailed Derivation of \eqref{eq:diff-1}} \label{app:derivation}
In this section we present a detailed derivation of \eqref{eq:diff-1}. The derivation strategy closely largely follows with \cite{Lin:2022rbf}, emphasizing precise normalization of states and keeping track of all approximations to ensure their validity. Due to the left/right symmetry, in the following derivation we shall only focus on the left Hamiltonian $H_L$. 

The one-particle spectrum is fully specified by the action of $H_{L/R}$ on the states. The left Hamiltonian can be represented in terms of $q$-ladder operators as:
\be
H_{L}=a_{L}+a_{L}^{\da},
\ee
where $a_L a_{L}^\da - q a_{L}^\da a_L = 1$. Its action on a one-particle state is used as defining property of wavefunctions in \ref{app:full-solution}. We have:
\be \label{eq:Hl-recur}
H_L|n_L,n_R\ra
=|n_L+1,n_R\ra+[n_L]|n_L-1,n_R\ra+q^{\Delta+n_L}[n_R]|n_L,n_R-1\ra,
\ee
where we have set $r_V=q^\Delta$. The matrix elements of $H_L$ should be defined in normalized states. In $0$-particle case this is simple to implement as states with different number chords are orthogonal. In $1$-particle case, however, since the overlap between states $|n_L,n_R\ra$ with equal $n_L +n_R$ is complicated, we do need to be careful when evaluating the matrix elements of $H_L$ in such basis. As an illustration, let us consider the overlap between $H_L |n_L , n_R \ra$ with $|n_L+1,n_R\ra$. We find that 
\be
\frac{\la n_{1}+1,n_{2}|H_{L}|n_{1},n_{2}\ra}{\sqrt{\la n_{1},n_{2}|n_{1},n_{2}\ra\la n_{1}+1,n_{2}|n_{1}+1,n_{2}\ra}}=\sqrt{\frac{\la n_{1}+1,n_{2}|n_{1}+1,n_{2}\ra}{\la n_{1},n_{2}|n_{1},n_{2}\ra}}.
\ee
The numerator can be evaluated by the recursive definition of inner product \eqref{eq:inner-product} as:
\be
\la n_{L}+1,n_{R}|n_{L}+1,n_{R}\ra=[n_{L}+1]\la n_{L},n_{R}|n_{L},n_{R}\ra+q^{n_{L}+\Delta+1}[n_{R}]\la n_{L},n_{R}|n_{L}+1,n_{R}-1\ra.
\ee
In the triple scaling limit we take both $n_L$ and $n_R$ to infinity, in this limit therefore the overlap between state $| n_L, n_R\ra$ and $|n_L+1, n_R -1\ra$ becomes equal to the overlap of $|n_L, n_R\ra$ with itself. Therefore, we have
\be \label{eq:approx-largen}
\la n_L, n_R \mid n_L+1, n_R-1 \ra \simeq \la n_L, n_R | n_L, n_R \ra.
\ee
Therefore, we find 
\be \label{eq:term1}
\frac{\la n_{1}+1,n_{2}|H_{L}|n_{1},n_{2}\ra}{\sqrt{\la n_{1},n_{2}|n_{1},n_{2}\ra\la n_{1}+1,n_{2}|n_{1}+1,n_{2}\ra}} \simeq \sqrt{\frac{1-\left(1-q^{\Del}\right)q^{n_{L}+1}-q^{n_{L}+n_{R}+\Del+1}}{1-q}}.
\ee
Similar strategy applies to the overlap with states in second and third term in \eqref{eq:Hl-recur}:
\be
\begin{split}
    &\frac{\left\langle n_L-1, n_R\left|H_L\right| n_L, n_R\right\rangle}{\sqrt{\left\langle n_L, n_R \mid n_L, n_R\right\rangle\left\langle n_L-1, n_R \mid n_L-1, n_R\right\rangle}}=\left[n_L\right] \sqrt{\frac{\left\langle n_L-1, n_R \mid n_L-1, n_R\right\rangle}{\left\langle n_L, n_R \mid n_L, n_R\right\rangle}} \\
    & \frac{\left\langle n_L, n_R-1\left|H_L\right| n_L, n_R\right\rangle}{\sqrt{\left\langle n_L, n_R \mid n_L, n_R\right\rangle\left\langle n_L, n_R-1 \mid n_L, n_R-1\right\rangle}}=q^{n_L+\Delta}\left[n_R\right] \sqrt{\frac{\left\langle n_L, n_R-1 \mid n_L, n_R-1\right\rangle}{\left\langle n_L, n_R \mid n_L, n_R\right\rangle}}.
\end{split}
\ee
Applying similar approximation as \eqref{eq:approx-largen}, we find 
\be
\begin{split} \label{eq:term2}
    \left[n_{L}\right]\sqrt{\frac{\left\langle n_{L}-1,n_{R}\mid n_{L}-1,n_{R}\right\rangle }{\left\langle n_{L},n_{R}\mid n_{L},n_{R}\right\rangle }}\simeq\sqrt{\frac{1-q^{n_{L}}}{1-q}}\times\left(1-\left(1-q^{\Del}\right)q^{n_{L}}-q^{n_{L}+n_{R}+\Del}\right)^{-1/2},
\end{split}
\ee
and 
\be
\begin{split} \label{eq:term3}
    q^{n_{L}+\Delta}\left[n_{R}\right]\sqrt{\frac{\left\langle n_{L},n_{R}-1\mid n_{L},n_{R}-1\right\rangle }{\left\langle n_{L},n_{R}\mid n_{L},n_{R}\right\rangle }}=& \sqrt{\frac{q^{n_{L}+\Del}-q^{n_{R}+n_{L}+\Del}}{1-q}}\\
    &\times\left(1-\left(1-q^{\Del}\right)q^{n_{L}}-q^{n_{L}+n_{R}+\Del}\right)^{-1/2}.
\end{split}
\ee
Now let us consider the triple scaling limit, where we introduce the renormalized length $l_{L/R}$ as
\be
q^{n_{L/R}}=\lambda e^{-\tilde{l}_{L/R}}.
\ee
The states $\lambda |n_L, n_R \ra$ can now be labeled with the length parameter, and we have
\be
|n_{L}+1,n_{R}\ra\simeq e^{\lambda\partial_{L}}|\tilde{l}_{L},\tilde{l}_{R}\ra.
\ee
We introduce the distance function $d(\tilde{l}_L , \tilde{l}_R)$ as:
\be
\ml\left(\tilde{l}_{L},\tilde{l}_{R}\right)=\lambda\left(1-e^{-\lambda\Del}\right)e^{-\tilde{l}_{L}}+\lambda^{2}e^{-\tilde{l}_{L}-\tilde{l}_{R}-\lambda\Del}.
\ee
Note that in $\Delta\to 0$ limit it produces the Liouville potential with total length $\tilde{l}=\tilde{l}_L +\tilde{l}_R$. Combining \eqref{eq:term1}, \eqref{eq:term2} and \eqref{eq:term3}, we find the Hamiltonian can be represented in terms of the new parameters as
\be \label{eq:Hl-new1}
\begin{split}
\tilde{H}_L = \lambda^{-1/2} H_L & =-\left(\sqrt{\frac{1-\ml\left(\tilde{l}_L, \tilde{l}_R\right)}{(1-q)\lambda}} e^{\lambda \tilde{\partial}_L}+e^{\lambda \tilde{\partial}_L} \sqrt{\frac{1-\ml\left(\tilde{l}_L, \tilde{l}_R\right)}{(1-q)\lambda}}\right) \\
& -\frac{1}{\sqrt{(1-q)\lambda}} \times \sqrt{\frac{e^{-\tilde{l}_L-\lambda \Delta}-e^{-\tilde{l}_L-\tilde{l}_R-\lambda \Delta}}{1-\ml\left(\tilde{l}_L, \tilde{l}_R\right)}}\left(e^{\lambda \tilde{\partial}_R}-e^{\lambda \tilde{\partial}_L}\right),
\end{split}
\ee
where we have rescaled $H_L$ by a factor of $\lambda^{1/2}$. The leading order of the right hand side in \eqref{eq:Hl-new1} gives a constant, which we denoted as $E_0$. Then by keeping terms up to $O(\lambda)$, we find 
\be
\tilde{H}_L-E_0=\lambda\left(-\tilde{\partial}_{\mathrm{L}}^{2}+e^{-\tilde{l}_{L}}\left(\Del+\partial_{R}-\partial_{L}\right)+e^{-\tilde{\ell}_{\mathrm{L}}-\tilde{\ell}_{\mathrm{R}}}\right) + O(\lambda^2).
\ee 
This matches up to a constant normalization with the Hamiltonian in the first equation in \eqref{eq:diff-1}.

\bibliographystyle{JHEP}
\bibliography{ref.bib}

\end{document}